\documentclass[pre,aps,preprint,showpacs,superscriptaddress]{revtex4-1}
\usepackage[T1]{fontenc}
\usepackage{revsymb4-1,latexsym,amssymb,amsmath,comment}
\usepackage{graphicx,psfrag,subfigure,stmaryrd,xcolor}
\usepackage[ansinew]{inputenc}
\usepackage{tabularx}
\usepackage{marvosym}
\usepackage{mathptmx}

\graphicspath{{./FIGS/}}

\newcommand{\be}{\begin{equation}}
\newcommand{\ee}{\end{equation}}
\newcommand{\bea}{\begin{eqnarray}}
\newcommand{\eea}{\end{eqnarray}}
\newcommand{\ba}{\begin{array}}
\newcommand{\ea}{\end{array}}
\newcommand{\bef}{\begin{figure}}
\newcommand{\ef}{\end{figure}}

\begin{document}

\author{Andreagiovanni Reina}
\email{a.reina@sheffield.ac.uk}
\affiliation{Department of Computer Science, University of Sheffield, UK}
\author{James A. R. Marshall}
\affiliation{Department of Computer Science, University of Sheffield, UK}
\author{Vito Trianni}
\affiliation{ISTC, Italian National Research Council, Rome, Italy}
\author{Thomas Bose}
\affiliation{Department of Computer Science, University of Sheffield, UK}

\title{Model of the best-of-$N$ nest-site selection process in honeybees}
\date{\today}

\begin{abstract}
  The ability of a honeybee swarm to select the best nest site plays a
  fundamental role in determining the future colony's fitness. To
  date, the nest-site selection process has mostly been modelled and
  theoretically analysed for the case of binary decisions. However,
  when the number of alternative nests is larger than two, the
  decision process dynamics qualitatively change. In this work, we
  extend previous analyses of a value-sensitive decision-making
  mechanism to a decision process among N nests. First, we present the
  decision-making dynamics in the symmetric case of N equal-quality
  nests. Then, we generalise our findings to a best-of-N decision
  scenario with one superior nest and N -- 1 inferior nests,
  previously studied empirically in bees and ants. Whereas previous
  binary models highlighted the crucial role of inhibitory
  stop-signalling, the key parameter in our new analysis is the
  relative time invested by swarm members in individual discovery and
  in signalling behaviours. Our new analysis reveals conflicting
  pressures on this ratio in symmetric and best-of-N decisions, which
  could be solved through a time-dependent signalling
  strategy. Additionally, our analysis suggests how ecological factors
  determining the density of suitable nest sites may have led to
  selective pressures for an optimal stable signalling ratio.

\pacs{87.23.Cc, 87.10.Ed, 87.23.Ge}

\end{abstract}

\maketitle

\section{Introduction}
Collective consensus decision-making \cite{Bose:COBS:2017}, in which
all members of a group must achieve agreement on which of several
options the group will select, is a ubiquitous problem. While groups
may be subject to conflicts of interest between members
(\textit{e.g.}~\cite{Conradt:TrendsEcolEvol:20:449:2005,
  Couzin:Science:334:1578:2011}), in groups where individuals'
interests align it is possible to look for mechanisms that optimise
group-level decisions \cite{Marshall:AAAI:12:2011}.  In this paper we
model collective consensus decision-making by social insect colonies,
in the form of house-hunting by honeybee swarms
\cite{Seeley:Science:2012, Pais:pone:8:e73216:2013}, but similar
decision-making problems manifest in diverse other situations, from
societies of microbes \cite{RossGillespie:fmicb:5:54:1:2014} to
committees of medical experts \cite{Kurvers:jamadermatol:15:1346:2015,
  Wolf:pone:10:0134269:2015}.
Much attention has been paid to optimisation of speed-accuracy
trade-offs in such situations
(\textit{e.g.}~\cite{Franks:PRSB:270:2457:2003,
  Marshall:JRSInt:3:243:2006, Marshall:JRSInterface:6:1065:2009,
  Pratt:PNAS:103:15906:2006, Golman:ScienceAdv:1:1500253}) but theory
shows that where decisions makers are rewarded by the value of the
option they select, rather than simply whether or not it was the best
available, managing speed-accuracy trade-offs may not help to optimise
overall decision quality \cite{Pirrone:FrontNeurosc:8:2014}. Here we analyse a value-sensitive
decision-mechanism inspired by cross-inhibition in house-hunting
honeybee swarms \cite{Seeley:Science:2012, Pais:pone:8:e73216:2013}.
One instance of value-sensitivity is the ability to make a choice when
the option value is sufficiently high---i.e., it exceeds a given
threshold. In case no option is available with high-enough value, the
decision maker may refrain from commitment to any option, in the
expectation that a high-quality option may later become available. As
a consequence, value-sensitivity is relevant above all in scenarios in
which multiple alternatives exist and possibly become available at
different times.  Another interesting property of the investigated
decision-making mechanism is its ability to break decision deadlocks
when the available options have equal quality. Deadlock breaking has
been shown to be of interest in a series of scenarios, including
collective motion \cite{Li:PRE:2008, Valentini:SI:2015}, spatial
aggregation \cite{Halloy:Science:2007, Hamann:NCA:2012} and collective
transport \cite{McCreery:pone:2016}.
Previous studies of value-sensitive decision-making have been limited
to binary decision problems, although it is known that honeybee swarms
and other social insect groups are able to choose from among many more
options during the course of a single
decision~\cite{Lindauer:Schwarmbienen:1955, Seeley:HoneybeeDemocracy,
  Seeley:BehavEcolSociobiol:2001, Franks:animalBehav:2006,
  Robinson:PONE:2011}. Here, we generalise the model of
\cite{Pais:pone:8:e73216:2013} and examine its ability to exhibit
value-sensitive deadlock-breaking when choosing between N equal
alternatives, and also to solve the best-of-N decision problem in
which one superior option must be selected over N -- 1 equal but
inferior distractor options.

\section{Mathematical model}

\subsection{General N-options case}

Our work builds on a previous model that describes the decentralised
process of nest-site selection in honeybee
swarms~\cite{Seeley:Science:2012}. The decentralised decision-making
process is modelled as a competition to reach threshold between
subpopulations of scout bees committed to an option (i.e., a
nest). The model is described as a system of coupled 
ordinary differential equations (ODEs), with each equation
representing the subpopulation committed to one option; an equation
describing how the subpopulation of uncommitted scout bees changes
over time is implicit, since the total number of bees in the system is
constant over the course of a decision.  Uncommitted scout bees
explore the environment and, when they discover an option $i$,
estimate its quality $v_i$, and may commit to that option at a rate
$\gamma_i$. The commitment rate to option $i$ for discovery is assumed
to be proportional to the option's quality, that is, more frequent
commitments to better quality nests ($\gamma_i \propto v_i$).
Committed bees may spontaneously revert, through abandonment, to an
uncommitted state at rate $\alpha_i$. Here, the abandonment rate is
assumed to be inversely proportional to the option's quality, that is,
poorer options are discarded faster ($\alpha_i \propto v_i^{-1}$).
This abandonment process allows bees quickly to discard bad options,
and endows the swarm with a degree of flexibility since bees are not
locked into their commitment state.
In addition to these two individual transitions, which we label as
\textit{spontaneous}, scout bees interact with each other to achieve
agreement on one option. In particular, the model proposed
in~\cite{Seeley:Science:2012} identifies two interaction forms:
recruitment and cross-inhibition, which give rise to
\textit{interaction} transitions. Recruitment is a form of positive
feedback, by which committed bees actively recruit, through the waggle
dance, uncommitted bees~\cite{Lindauer:Schwarmbienen:1955,
  VonFrisch:Book:Bees:1967, Seeley:BehavEcolSociobiol:45:19:1999}.
Therefore, the rate by which uncommitted bees are recruited to option
$i$ is determined by both the number of bees committed to $i$ and the
strength of the recruitment process for $i$, labelled as $\rho_i$.
Similarly to discovery, recruitment is assumed to be proportional to
the option's quality ($\rho_i \propto v_i$). The other interaction
form that occurs in this decision process is cross-inhibition.
Cross-inhibition is a negative feedback interaction between bees
committed to different options; when a bee committed to option $i$
encounters another bee committed to another option $j$, (with
$j\neq i$), the first may deliver stop signals to the second which
reverts to an uncommitted state at a rate $\beta_{ij}$. For binary
choices stop-signalling has previously been shown to be a control
parameter in a value-sensitive decision-making mechanism, in
particular setting a value threshold for deadlock maintenance or
breaking in the case of equal-quality
options~\cite{Seeley:Science:2012, Pais:pone:8:e73216:2013}.
In this study, in agreement with the assumptions made above, we assume
cross-inhibition proportional to the quality of the option that the
bees delivering the stop signal are committed to. In other words, bees
committed to better options will more frequently inhibit bees
committed to other options ($\beta_{ij} \propto v_i$, see
Section~\ref{sec:new-param} for more details).

As described above, the set of bees committed to the same option is
considered as a sub-population, and the model describes changes in the
proportion of bees in each sub-population with respect to the whole
bee population. We assume that a decision is reached when one decision
sub-population reaches a quorum threshold \cite{Pratt:BES:52:117:2002,
  Sumpter:PTRSB:364:743:2008,
  Seeley:BehavEcolSociobiol:56:594:2004}. Precisely, $x_i$ and $x_u$
denote the proportion of bees committed to option $i$ and uncommitted
bees, respectively, with $N$ options and $i \in \{1,\dots,N\}$. A
version of the model that we analyse in this study has been originally
proposed for the binary decision case (i.e., $N=2$)
in~\cite{Seeley:Science:2012} and, later, extended to a more general
case of $N$ options in~\cite{Reina:10:pone:0140950:2015}. Analysis of
the value-sensitive parameterisation has been presented by Pais et
al. in~\cite{Pais:pone:8:e73216:2013}. Here we generalise this model,
and extend its analysis to the best-of-N case. The general models is:

\begin{align} 
  \label{EqSysOrigParam}
  \left\{ 
    \begin{aligned}
      \frac{dx_i}{dt} &= \gamma_i\, x_u - \alpha_i\, x_i + \rho_i\, x_u\, x_i 
                        - \sum_{j=1}^N\, x_j\, \beta_{ji}\, x_i \,,\qquad  i \in \{1,\,\dots\,,\,N\} \,, \\
      x_u &= 1 -  \sum_{i=1}^N x_i 
    \end{aligned}
  \right.
\end{align}

\subsection{A novel parameterisation for value-sensitive decision-making}
\label{sec:new-param}

Following earlier work~\cite{Seeley:Science:2012,
  Pais:pone:8:e73216:2013, Marshall:JRSInterface:6:1065:2009}, we
assume a value-sensitive parameterisation by which the transition
rates are proportional (or inversely proportional) to the option's
quality $v_i$, as mentioned above. Previous work investigated the
dynamics of the system~\eqref{EqSysOrigParam} with
\mbox{$v_i = \gamma_i=\rho_i=\alpha_i^{-1}$} and $\beta_{ij} = \beta$
for two options (i.e., $N=2$) \cite{Pais:pone:8:e73216:2013}. Such a
parameterisation displays properties that are both biologically
significant, and of interest for the engineering of artificial swarm
systems~\cite{Reina:10:pone:0140950:2015,
  Reina:SwarmIntelligence:9:75:2015}. One of the main system
characteristics is its ability to adaptively break or maintain
decision deadlocks when choosing between equal-quality options, as a
function of those options' quality. In fact, it has been shown that
when the swarm has to decide between two equally and sufficiently good
options, it is able to implement the best strategy: that is, to
randomly select any of the two options in a short time. However, in
Appendix~\ref{sec:pais-deadlock} we show that the system's dynamics
qualitatively change for more than two options, i.e., $N>2$: by
adopting the parameterisation proposed
in~\cite{Pais:pone:8:e73216:2013}, the swarm cannot break a decision
deadlock for more than two equally good options (see
Figure~\ref{fig:Bifurcation3DforN2} and
Appendix~\ref{sec:pais-deadlock}).

In this study, we extend previous work by introducing a novel
parameterisation that features value-sensitivity also for $N>2$.
Unlike~\cite{Pais:pone:8:e73216:2013}, we investigate a more general
parameterisation in which we decouple the rates of spontaneous
transitions (i.e., discovery and abandonment) from the rates of
interaction transitions (i.e., recruitment and cross-inhibition),
similarly to~\cite{Reina:10:pone:0140950:2015}. The proposed
parameterisation is $\gamma_i=k\,v_i$, $\alpha_i=k/v_i$ and
$\rho_i=h\,v_i$, where $k$ and $h$ modulate the strength of
spontaneous and interaction transitions, respectively.

For the cross-inhibition parameter, we consider the general case in
which $\beta_{ij}$ is the product of two components:
$\beta_{ij}=[A\cdot D]_{ij}$, where $A$ and $D$ are two matrices and
$\beta_{ij}$ is the $ij^{th}$ element of their product. The former,
$A$, is an adjacency matrix that expresses how subpopulations interact
with each other. Therefore, the entries $a_{ij}$ of $A$ are either $1$
or $0$ depending on whether interactions between subpopulations $i$
and $j$ can occur or not. The introduction of the adjacency matrix
allows us to define if inhibitory messages are delivered only between
bees committed to different options (i.e., cross-inhibition), or also
between bees committed to the same option (i.e., self-inhibition, as
\textit{self} refers to the own subpopulation). In this study, in
accordance with behavioural results in the
literature~\cite{Seeley:Science:2012}, we do not include
self-inhibitory mechanisms; thus the adjacency matrix contains zeros
along its diagonal (i.e., $a_{ii}=0, \forall i$). On the other hand,
we consider that interactions between different subpopulations are
equally likely, and this is reflected by having
$a_{ij} = 1, \forall i \neq j$. The second component, $D$, is a matrix
that quantifies the stop-signal strength, and allows us to define, if
needed, different inhibition strengths for each sender/receiver
couple. In other words, through $D$ the inhibitory signals can be
tuned not only as a function of the option quality of the inhibiting
population, but also as a function of the option quality of the
inhibited population.
In this analysis, we model dependence of cross-inhibition strength
solely on the value of the option that inhibiting bees are informed
about; thus we investigate the system dynamics for 
a diagonal cross-inhibition matrix with values $h\,v_1, \dots, h\,v_N$
along its diagonal, where $h$ is a constant interaction term (as for
recruitment), and
the $v_i, i\in\{1,\dots,N\},$ are qualities of the options the
inhibiting populations are committed to. Hence we parameterise the
cross-inhibition term as $\beta_{ij}=A_{ik}D_{kj}=hv_i$, which
determines the other parameters of the system
as~\eqref{EqSysOrigParam}

\begin{equation}
  \gamma_i = k \, v_i, \qquad
  \alpha_i = k \, v_i^{-1}, \qquad
  \rho_i = h \, v_i, \qquad
  \beta_{ij} = h \, v_i \,.
  \label{eq:params}
\end{equation}

In the following, we introduce the ratio $r=h/k$ between interaction
and spontaneous transitions. The ratio $r$ acts as the control
parameter for the decision-making system under our new formulation,
whereas the strength of cross-inhibition (stop-signalling rate) was
the control parameter in the original
analysis~\cite{Pais:pone:8:e73216:2013}. This new control parameter
has a simple and natural biological interpretation, as the propensity
of scout bees to deliver signals to others (here, represented by the
interaction term $h$), relative to the rate of spontaneous transitions
(here, represented by the term $k$).

We show that the novel parameterisation displays the same
value-sensitive decision-making properties of the binary system that
are shown in previous studies~\cite{Pais:pone:8:e73216:2013}.
In particular, we confirm that, in the symmetric case of two
equal-quality options, the ratio of interaction/spontaneous
transitions, \mbox{$r=h/k$}, determines when the decision deadlock is
maintained or broken (see Figure~\ref{fig:N2symmetric}). Additionally,
we show in Figure~\ref{fig:weberN2} that the interaction ratio
$r$ determines the just-noticeable difference to discriminate between
two similar value options, in a manner similar to Weber's law, as
demonstrated for the cross-inhibition rate
in~\cite{Pais:pone:8:e73216:2013}.

\subsection{The best-of-N decision problem}

As well as presenting a general analysis of the system dynamics for
small $N$ ($N=3$), for larger values of $N$ we next analyse the
best-of-N decision scenario with one superior and $N-1$ inferior
options. This scenario is consistent with empirical studies undertaken
with bees~\cite{Seeley:BehavEcolSociobiol:2001},
ants~\cite{Franks:animalBehav:2006, Robinson:PONE:2011} and with
neurophysiological studies~\cite{Schall:nature:1993}.
Considering such a scenario allows us to investigate the system
dynamics as a function of four parameters: (i)~the number of options
$N$, (ii)~the superior option $s$'s quality $v=v_s$, (iii)~the ratio
between the quality of any of the equal-quality inferior options and
of the superior option $\kappa=v_i/v_s$ (with $i\neq s$), and (iv)~the
ratio between interaction and spontaneous transitions $r=h/k$.  The
system of Equation~\eqref{EqSysOrigParam} with the parameterisation
given in \eqref{eq:params} can be rewritten in terms of these four
parameters as:
\begin{align}
  \left\{ 
  \begin{aligned}
    \frac{dx_1}{d\tau} &= v\, x_u - \frac{x_1}{v} + r\,v\, x_1
    \left[x_u - \sum_{j\neq 1}\,\kappa\, x_j \right]
    \,,\\
    \frac{dx_i}{d\tau} &= v\, \kappa\, x_i - \frac{x_i}{v\, \kappa} +
    r\,v\, x_i \left[ \kappa ( x_u - \sum_{j\neq 1,i}\, x_j) - x_1
    \right] \,,
    \qquad  i = 2,\,...\,,\,N \,, \\
    x_u &= 1 - \sum_{i=1}^N x_i
  \end{aligned}
  \right.
  \label{eq:fullSysBestOfN}
\end{align}
where $x_1$ is the population committed to the best (superior) option
(i.e., $v=v_1 \ge v_i ,\, \forall i \in \{2,\dots,N\}$) and
$\tau = k\,t$ is the dimensionless time.

The system in~\eqref{eq:fullSysBestOfN} is characterised by $N$
coupled differential equations and one algebraic equation. In
Equations~\eqref{eq:reducedSys-simple}, we reduce this system to a system
of two coupled differential equations by aggregating the dynamics of
the populations committed to the inferior options. In
Section~\ref{results}, we show that this system reduction allows us to
attain qualitatively correct results for arbitrarily large
$N$.

\section{Results\label{results}}

We first investigate the system dynamics for the case of $N=3$
options, then we generalise our findings to arbitrarily large $N$.
The reduced system (Equation~\eqref{eq:reducedSys-simple}) allows us to
investigate the dynamics for arbitrarily large numbers of options $N$
without increasing the complexity of the analysis.  In
Section~\ref{sec:results-sym}, we show the analysis results for the
symmetric case of $N$ equally good options, while in
Section~\ref{sec:results-asym}, we report the results for different
quality options.

\subsection{Symmetric case}\label{sec:results-sym}

\begin{figure}
\centering
  \subfigure[]{\label{fig:stabilityN3sym}
\includegraphics[width=0.32\linewidth]{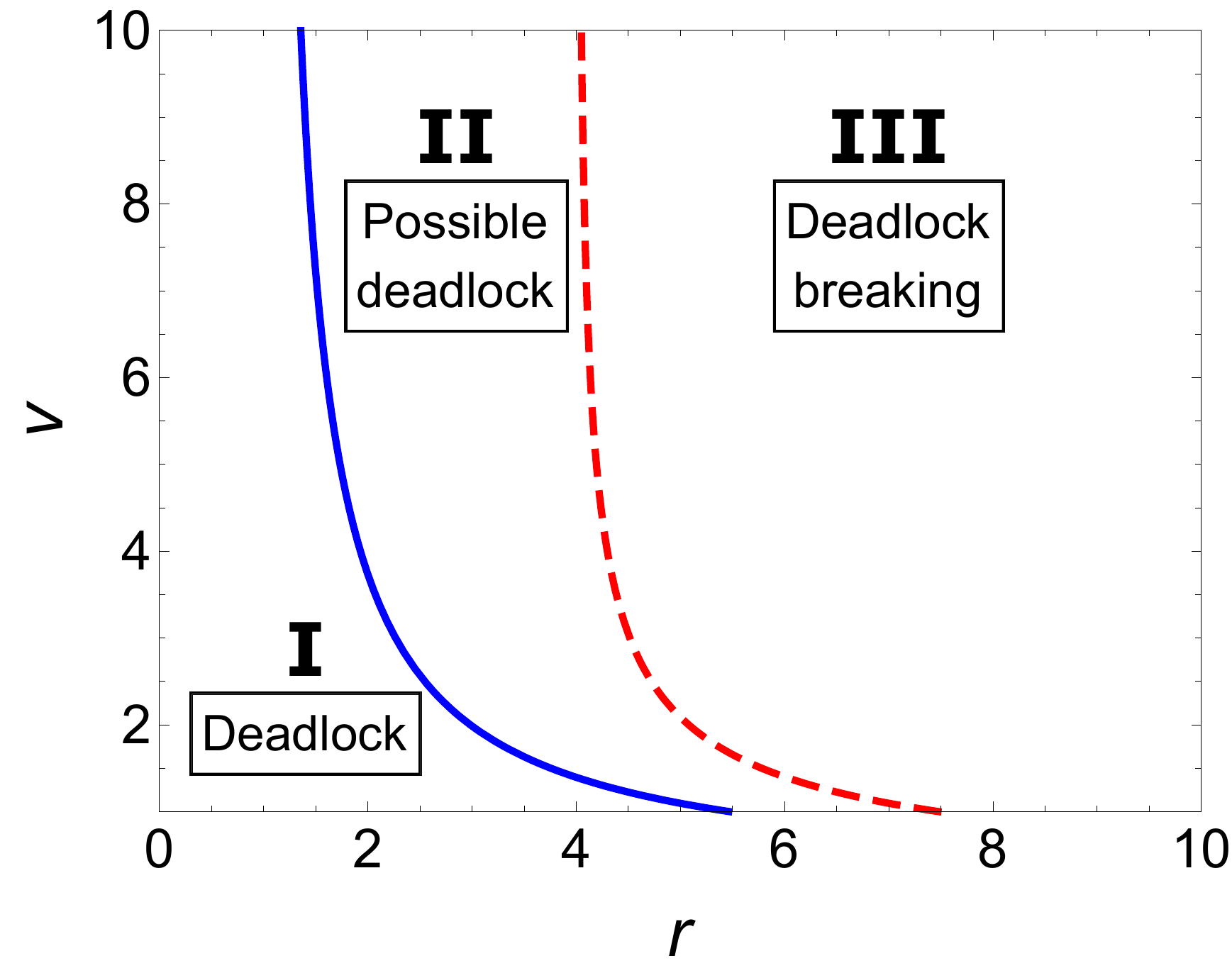}}
  \subfigure[]{\label{fig:bifurcationN3sym}
    \includegraphics[width=0.32\linewidth]{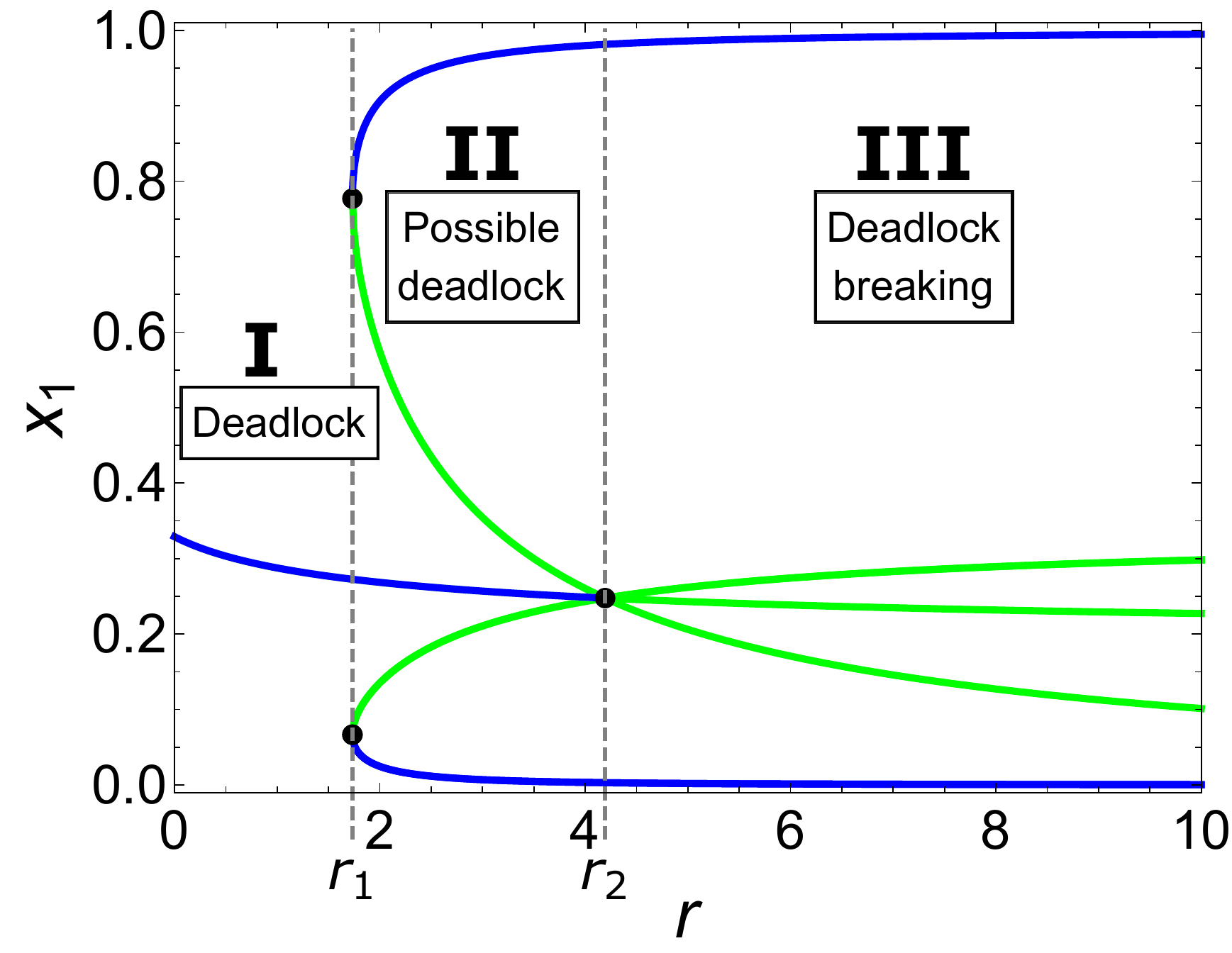}}
  \subfigure[]{\label{fig:bifOnNsym}
    \includegraphics[width=0.32\linewidth]{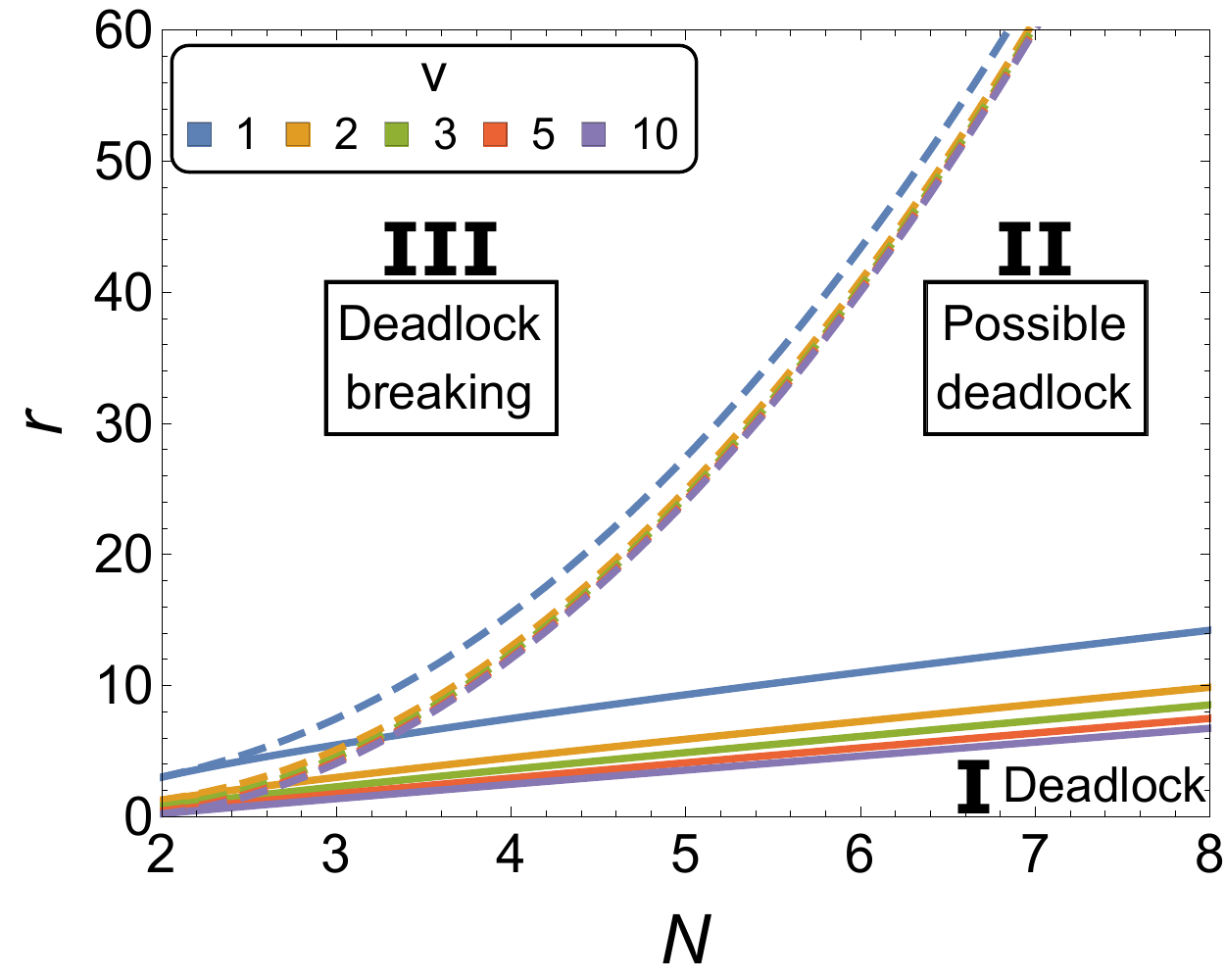}}
  \caption{Dynamics of the complete decision system of
    Equation~\eqref{eq:fullSysBestOfN} for the symmetric case
    $\kappa=1$ (i.e.,
    $v_1=v_2=v_3=v$). Panel~\subref{fig:stabilityN3sym} shows the
    stability diagram as a function of the parameter $r$ and the
    quality $v$ for $N=3$ options. The two curves represents the two
    bifurcations $r_1$ (blue solid) and $r_2$ (red dashed) of
    Equations~\eqref{eq:bifurcations-main}. There are three possible
    system phases: (I)~decision-deadlock, (II)~coexistence of decision
    deadlock and stable solutions for any option, and (III)~decision
    for any option. Panel~\subref{fig:bifurcationN3sym} shows the
    bifurcation diagram for $N=3$ and $v=5$ as a function of the
    parameter $r$. This illustrates the three system phases when
    varying the control parameter $r$. Note that, due to the 2D
    visualisation, some equilibria overlap and thus the bottom
    branches in panel~\subref{fig:bifurcationN3sym} correspond to the
    two overlapping equilibria for the options $x_2$ and
    $x_3$. Panel~\subref{fig:bifOnNsym} shows a stability diagram that
    visualises the dependence of the bifurcation points $r_1$ (solid
    lines) and $r_2$ (dashed lines) as a function of $N$ for varying
    $v \in \{1,2,3,5,10\}$, and reports the same three system phases.}
\label{fig:dynamicsN3sym}
\end{figure}

\begin{figure}
  \centering
  \subfigure[] {\label{fig:portrait-sym-I}
    \includegraphics[width=0.32\linewidth]{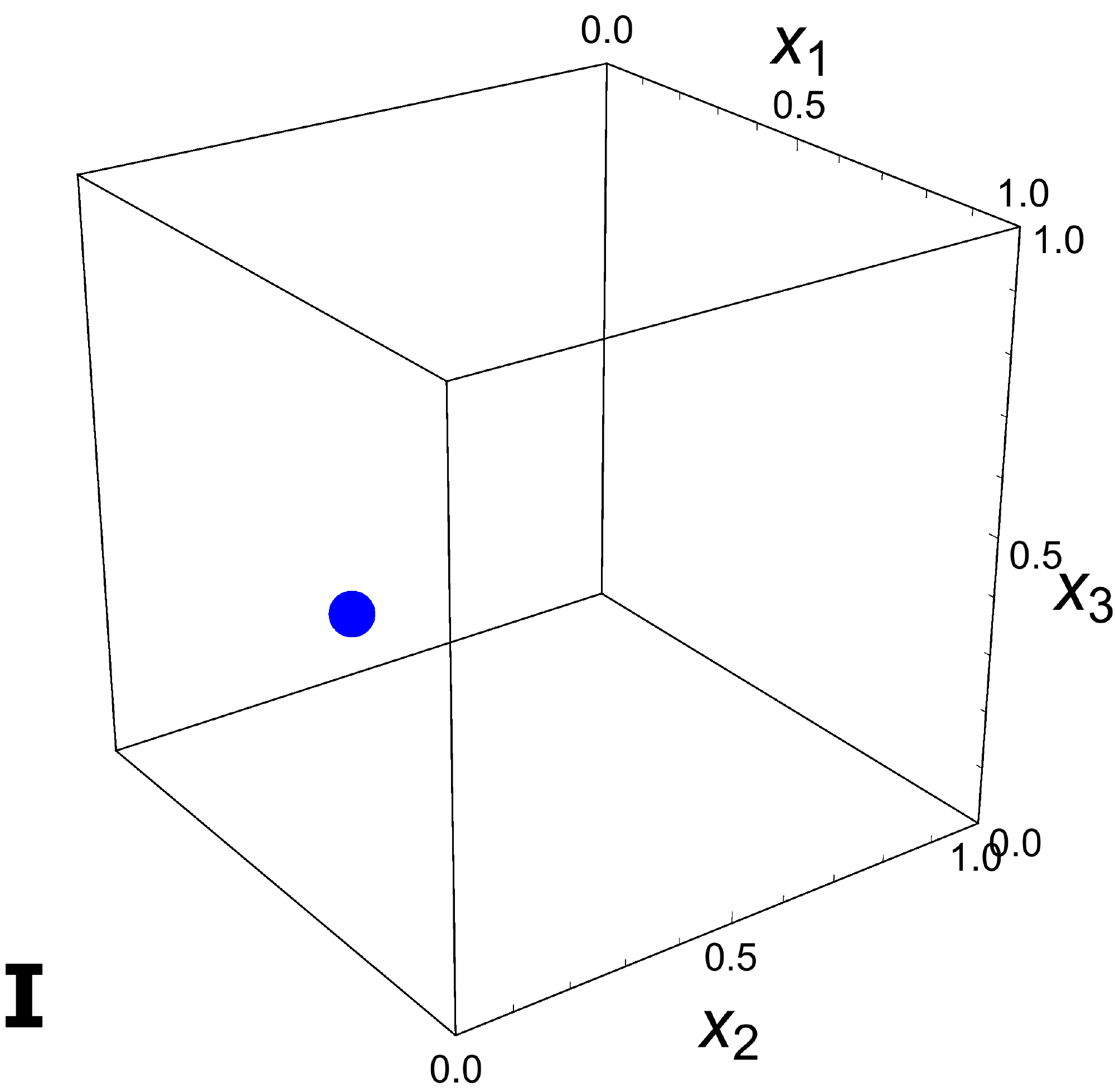}}
  \subfigure[] {\label{fig:portrait-sym-II}
    \includegraphics[width=0.32\linewidth]{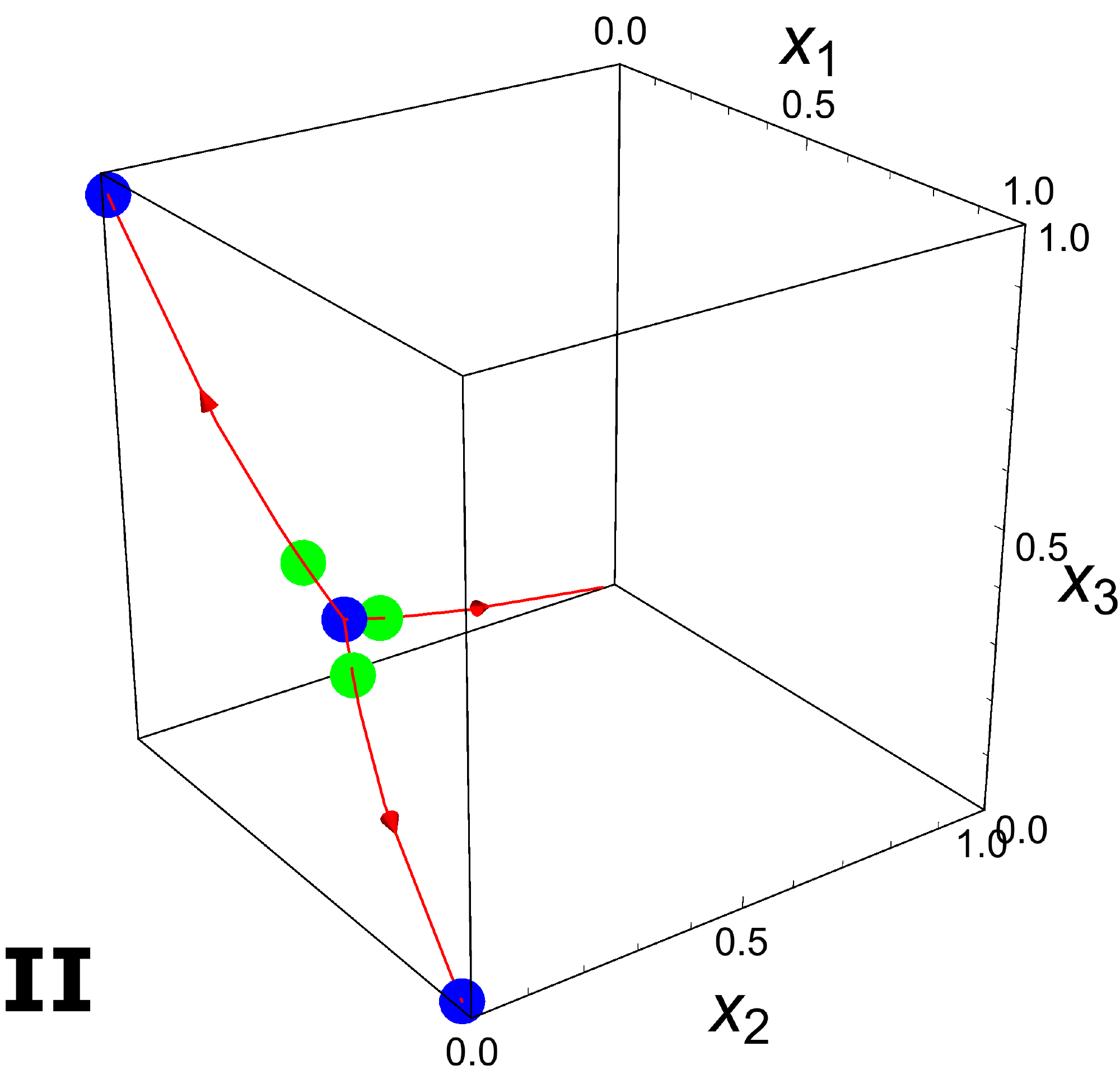}}
  \subfigure[] {\label{fig:portrait-sym-III}
    \includegraphics[width=0.32\linewidth]{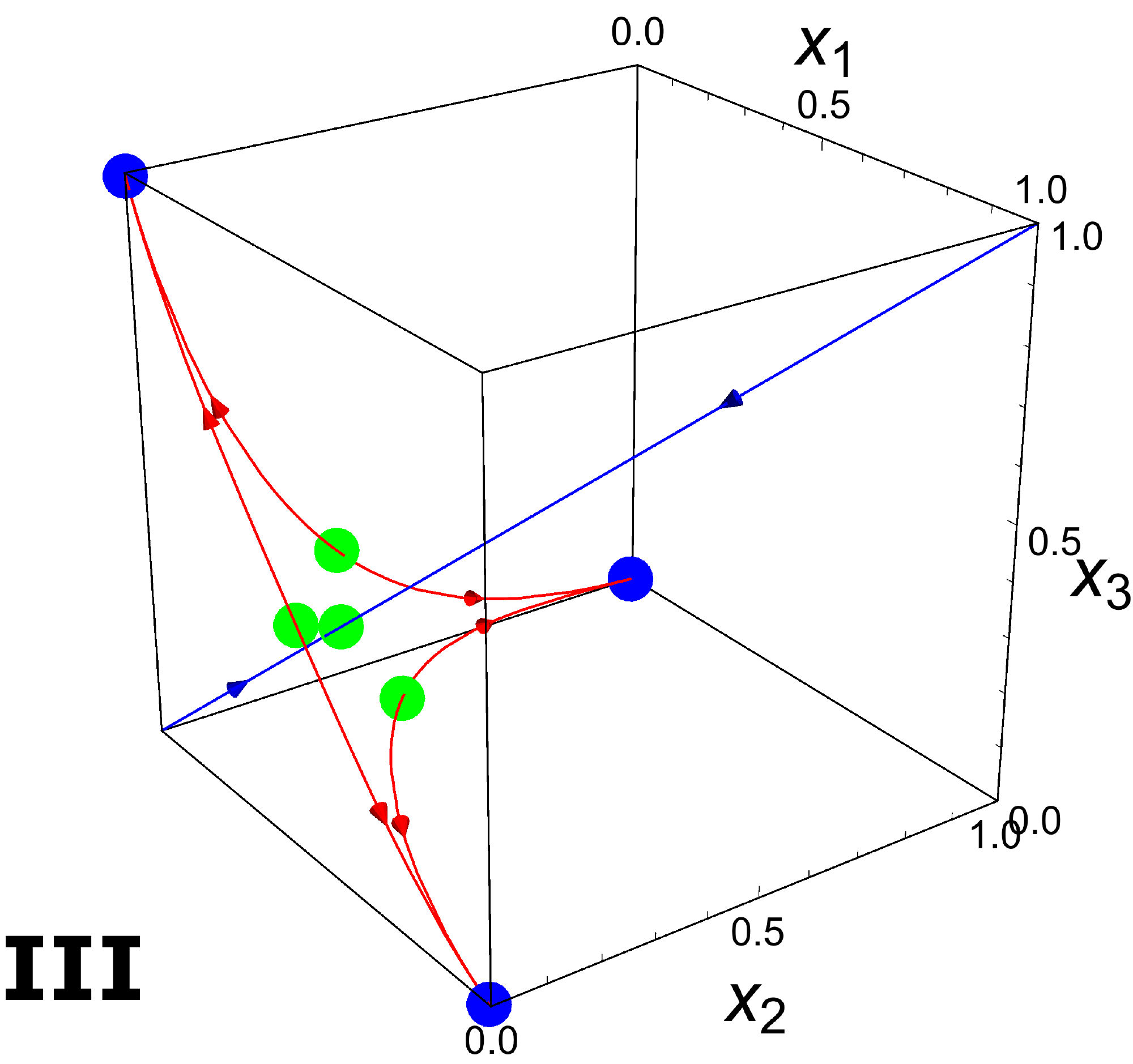}}
\begin{minipage}[b]{0.61\textwidth}
  \caption{Phase portraits of the complete
    system~\eqref{eq:fullSysBestOfN} for $N=3$ options in the
    symmetric case $\kappa=1$ (i.e., $v_1=v_2=v_3=v=5$). Blue dots
    represent stable equilibria, and green dots represent unstable
    saddle points. Saddle manifolds are shown as red (repulsive) and
    blue (attracting) lines. Panel~\subref{fig:portrait-sym-I} shows
    the system in a decision deadlock phase (i.e., phase I of
    Figure~\ref{fig:bifurcationN3sym}, $r=1$), in fact, there is only
    one stable solution with all the three committed population with
    equal size. Panel~\subref{fig:portrait-sym-II} shows the
    coexistence of the decision deadlock and the decision for any
    option (phase II, $r=3$). Panel~\subref{fig:portrait-sym-III}
    shows the system for high values of $r$ in which the decision
    deadlock solution is an unstable saddle point, and therefore the
    only stable solutions are the decision for any option (phase III,
    $r=10$). The same phase portrait from another perspective is shown
    in panel~(d) where a set of trajectories (red lines) are
    shown. Looking at panel~(d), the central unstable saddle node is
    unstable on the displayed plane while is stable (i.e., attracting)
    on the direction orthogonal to the field of view of the plot~(d)
    (i.e., the attraction manifold is the line
    $x_1=x_2=x_3$). The system does not possess any periodic
      attractors.}
  \label{fig:portrait-sym}
  \end{minipage}
  \hspace{0.03\linewidth}
\subfigure[] {
    \addtocounter{subfigure}{+3}
    \label{fig:portrait-cartoon}
    \includegraphics[trim={0 0 0 0cm}, width=0.32\linewidth]{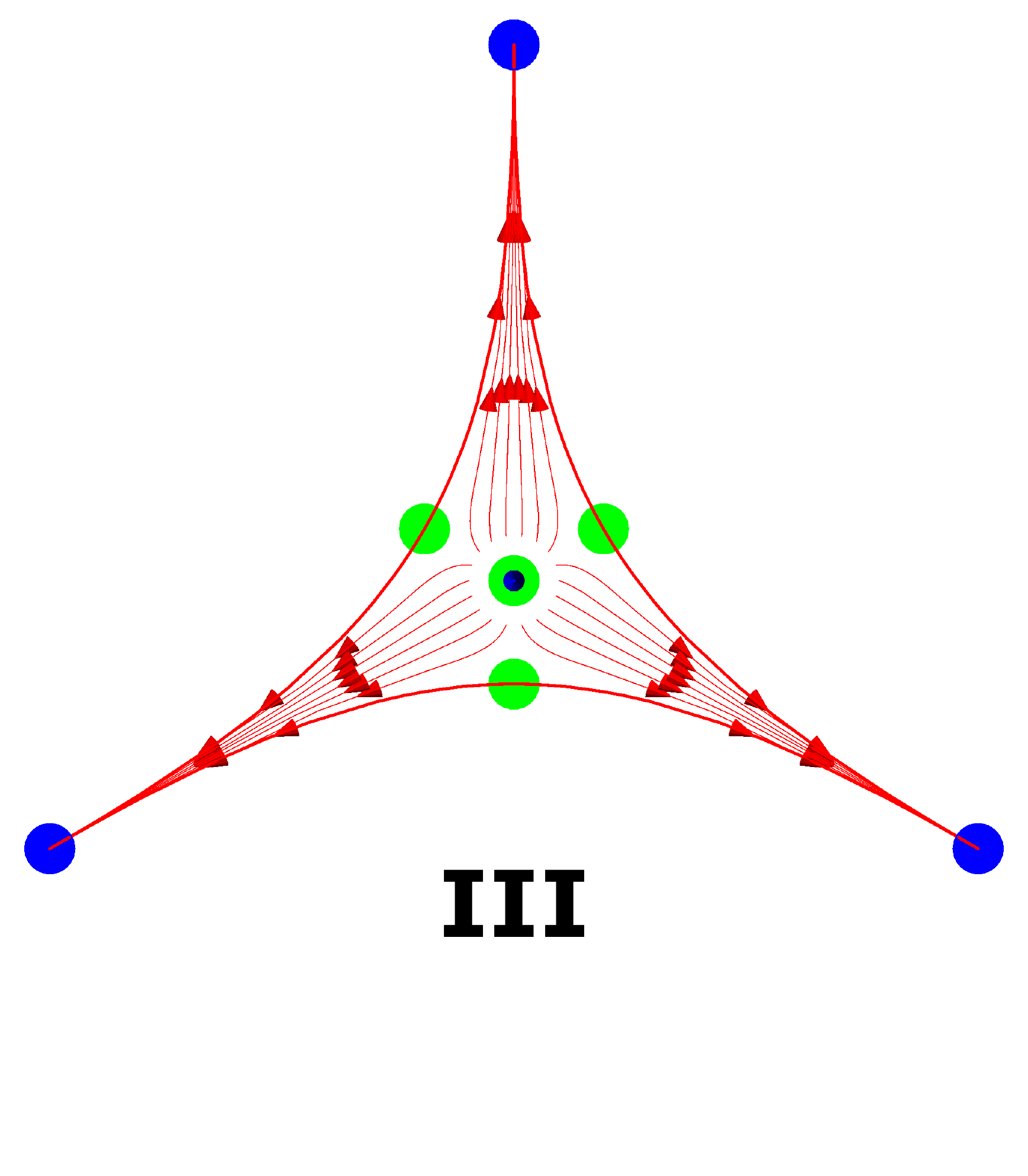}}
\end{figure}

We start by analysing the symmetric case of $N$ equal-quality options
(i.e. $\kappa=1$).  The simplicity of the reduced system
(Equation~\eqref{eq:reducedSys-simple}) allows us to determine the
existence of two bifurcation points which are determined by the
parameters $r$, $v$ and $N$, and we show the bifurcation conditions in
terms of the control parameter $r$ as:
\begin{equation}
  \label{eq:bifurcations-main}
    r_1 = f_1(v,N), \qquad 
    r_2 = f_2(v,N)\, .
\end{equation}
In Appendix~\ref{sec:bifurcations}, we report the complete equations
for \eqref{eq:bifurcations-main} as functions of $(v,N)$ (see
Equation~\eqref{eq:bifurcations}) or, more generally, of
$(\gamma$,$\alpha$,$\rho,\beta)$ (see
Equation~\eqref{eq:bifurcation-general}).
In Figure~\ref{fig:stabilityN3sym}, we show the stability diagram of
the system~\eqref{eq:fullSysBestOfN} in the parameter space $(r,v)$,
for $N=3$. When the pair $(r,v)$ is in area~I, the system cannot break
the decision deadlock but remains in an undecided state with an equal
number of bees in each of the three committed populations. This result
can be also seen in Figure~\ref{fig:bifurcationN3sym}, where we
display the bifurcation diagram for the specific case $v=5$. Here, low
values of $r$ correspond to a single stable equilibrium representing
the decision deadlock. Increasing the signalling ratio, the system
undergoes a saddle node bifurcation when $r=r_1$ in
Figure~\ref{fig:bifurcationN3sym}, at which point a stable solution
for each option appears and the selection by the swarm of any of the
$N$ equally-best quality options is a feasible solution. However, for
$(r,v)$ in area~II of Figure~\ref{fig:stabilityN3sym}, the
decision-deadlock remains a stable solution and only through a
sufficient bias towards one of the options the system converges
towards a decision. This system phase can be visualised in the
bifurcation diagram of Figure~\ref{fig:bifurcationN3sym} and in the
phase portrait of Figure~\ref{fig:portrait-sym-II}: The system escapes
from the decision-deadlock attraction basin if noise leads the
population to jump into a neighbouring basin corresponding to a unique
choice.

The system undergoes a second bifurcation at $r=r_2$ in
Figure~\ref{fig:bifurcationN3sym}, that changes the stability of the
decision-deadlock from stable ($r<r_2$) to partially unstable (saddle,
$r>r_2$). Therefore, for sufficiently high values of the signalling
ratio (area~III in Figure~\ref{fig:stabilityN3sym}), the unique
possible outcome is the decision for any of the equally best quality
options. The central solution of indecision remains stable (i.e.,
attracting) with respect to only one manifold, i.e., the line for
equal-size committed populations, while it is unstable with respect to
the other directions (see the phase portraits of
Figures~\ref{fig:portrait-sym-III}-\subref{fig:portrait-cartoon} and the
video in the supplemental material \cite{supp-video}).
Instead, the
unstable saddle points that surround the central solution have
opposite attraction/repulsion manifolds. For this reason, several
unstable equilibria can be near to each other, as in
Figure~\ref{fig:bifurcationN3sym}.

The analysis of the system with three options reveals three system
phases as a consequence of the two bifurcations determined by $f_1$
and $f_2$ (Equation~\eqref{eq:bifurcations-main}). Increasing the number of
options, the number of system phases increases. In particular, for
every other $N$, at odd values (i.e., $N\in\{5,7,9,\dots\}$), a new
bifurcation point between $r_1$ and $r_2$ appears. In
Figure~\ref{fig:bifurcationsManyNs}, we report the bifurcation
diagrams for $v=5$ and $N \in \{4,5,6,7\}$.  Despite the system phase
increase, the main dynamics for any $N>2$ can be described by the
three macro-phases described above: (I) decision-deadlock only, (II)
coexistence of decision-deadlock and decision, and (III) decision
only. In fact, the additional equilibria that appear for odd $N$ are
all unstable saddle solutions (with orthogonal attraction/repulsion
directions with each other) which do not change the stability of other
solutions. Therefore, we focus our study on the bifurcations defined
by Equations~\eqref{eq:bifurcations-main} (i.e., Eq.~\eqref{eq:bifurcations}) 
which determine the main phase
transitions.

Figure~\ref{fig:bifOnNsym} shows the relationship between the
bifurcation points $r_1$ and $r_2$, the options's quality $v$ and the
number of options $N$. The effect of $v$ on $r_1$ and $r_2$ remains
similar to that seen in Figure~\ref{fig:stabilityN3sym}, i.e., the
bifurcation points vary as a function of $v$ when $v$ is low, while
they are almost independent of $v$ when it is large. More precisely,
the influence of the quality magnitude $v$ on the system dynamics
decreases quadratically with $v$ (see
Equation~\eqref{eq:bifurcations}). The number of options, $N$,
influences differently the two bifurcation points. While $r_1$ grows
quasi-linearly with $N$, instead $r_2$ grows quadratically with $N$.
Therefore, in the symmetric case, the number of options that the swarm
considers plays a fundamental role in the decision dynamics. In fact,
too many options preclude the possibility of breaking the
decision-deadlock and selecting one of the equally-best options. This
result suggests a limit on the maximum number of equal options that
can be concurrently evaluated by the modelled decision-maker.

\subsection{Asymmetric case}\label{sec:results-asym}

\begin{figure}
\centering
  \subfigure{ 
    \includegraphics[width=0.675\linewidth]{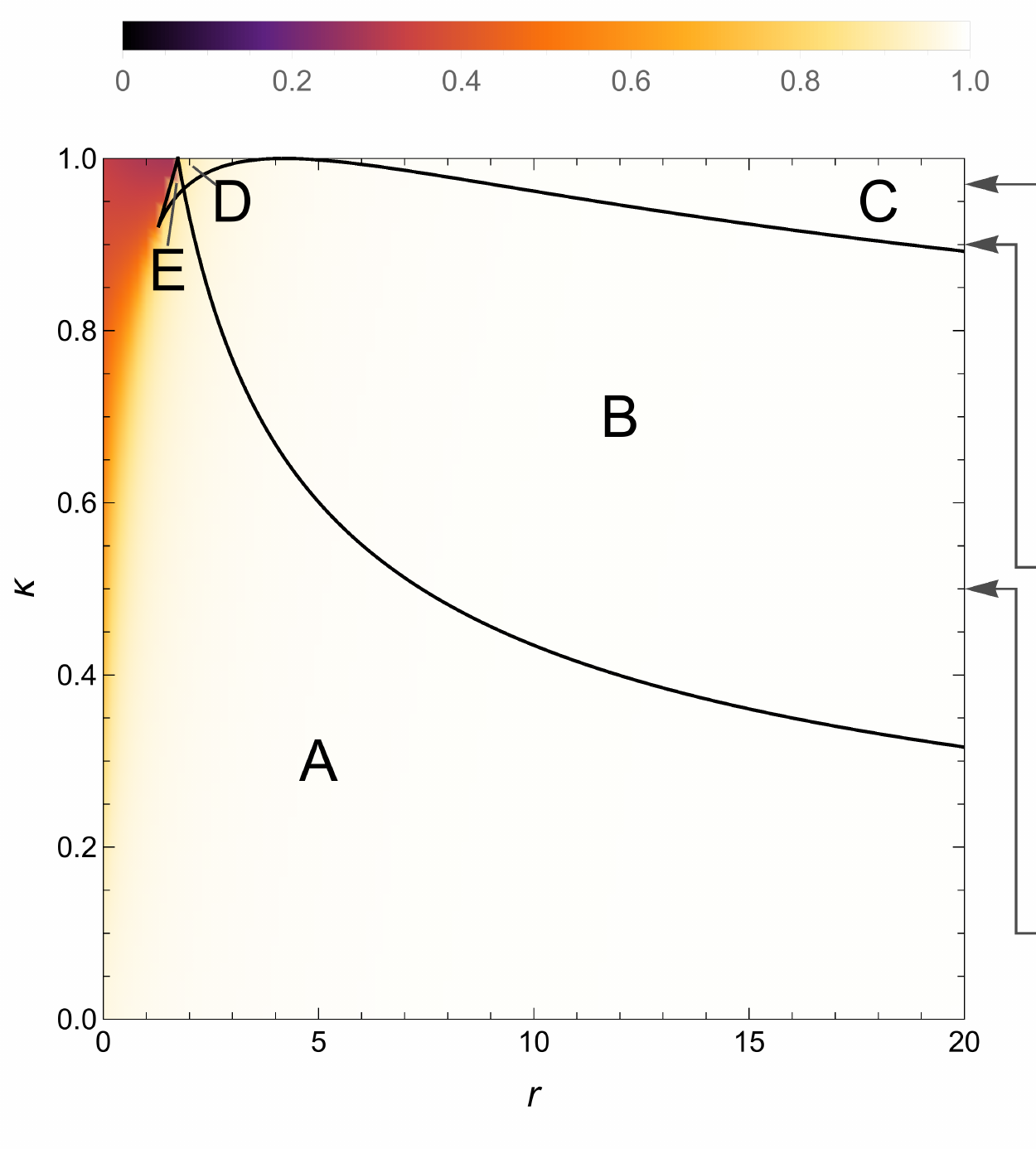}}
  \subfigure{ 
    \includegraphics[width=0.305\linewidth,trim=1cm 0 0 0]{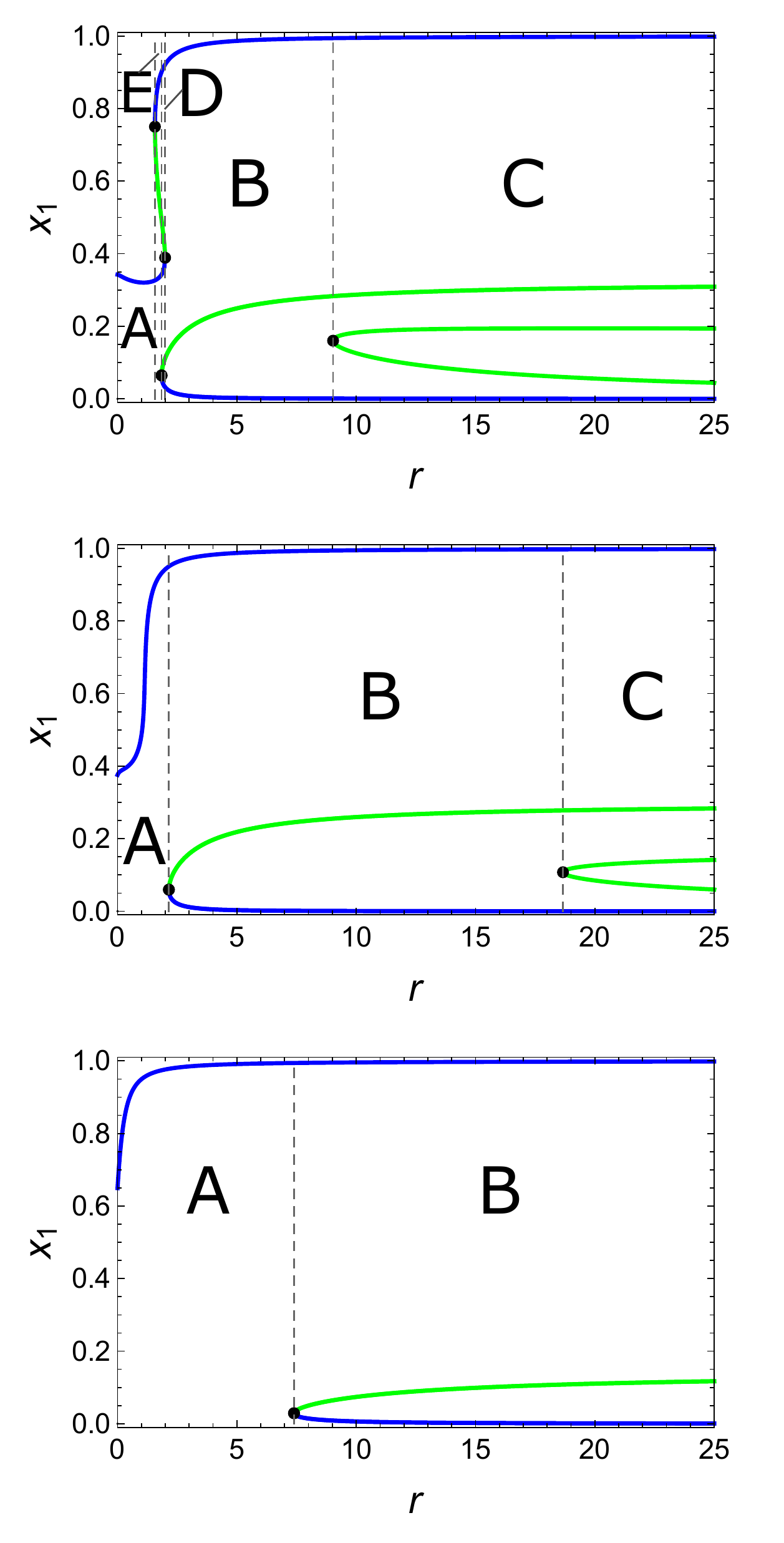}}
  \caption{Dynamics of the complete decision system of
    Equations~\eqref{eq:fullSysBestOfN} for $N=3$ options for the
    asymmetric case ($\kappa<1$) and superior option's quality
    $v=5$. The left panel shows the stability diagram as a function of
    the parameter $r$ and the ratio between qualities $\kappa$. The
    parameter space is divided in five different areas (see
    Figure~\ref{fig:stabilityRK} to see a representative 3D phase
    portrait for each area). In area A, the system has a unique
    solution corresponding to selection of the best option; in areas B
    and C, the system may select any of the possible options; in areas
    D and E the system may end in a decision deadlock. The underlying
    density map show the population size of the stable solution for
    the best option. For low values of $r$ and similar options
    (top-left corner), this population is relatively small and may be
    not enough to reach a quorum threshold. The right panels show
    three bifurcation diagrams as a function of the parameter $r$ for
    $\kappa \in \{0.5,0.9,0.97\}$.  Note that, due to the 2D
    visualisation, some equilibria overlap and thus the bottom
    branches of the bifurcation diagrams correspond to two overlapping
    equilibria for selection of options $x_2$ and $x_3$.}
\label{fig:dynamicsN3asym}
\end{figure}

We next analyse the system dynamics in the asymmetric best-of-N case
where option 1 is superior to the other $N-1$ same-quality,
inferior options $i$ (with $i \in \{2,\dots,N\}$).
Figure~\ref{fig:dynamicsN3asym} shows the stability diagram for $N=3$
options in the paremeter space $r,\kappa$. The results show that low
values of $r$ allow the system to have a unique solution, (area A in
the left panel of Figure~\ref{fig:dynamicsN3asym}). This is especially
true when the difference between the options is larger (i.e., low
values of $\kappa$). However, such stable solutions may not correspond
to a clear-cut decision, as the population fraction committed to the
best alternative may be too low to reach a decision threshold, as
indicated by the underlying density map in
Figure~\ref{fig:dynamicsN3asym}: if $r$ is small and $\kappa$
sufficiently high, only about half of the population will be committed
to the best option. Hence, a sufficiently high value of $r$ is
required for the implementation of a collective decision. For larger
values of $r$, the system undergoes various bifurcations leading to
$N$ stable solutions corresponding to the selection of each available
option (areas B and C of the left panel in
Figure~\ref{fig:dynamicsN3asym}). Therefore, there is the possibility
that an inferior option gets selected. For high values of $\kappa$,
two additional areas appear, labelled D and E in
Figure~\ref{fig:dynamicsN3asym}. These areas correspond to the
co-existence of an undecided state together with a decision state for
the superior and/or the inferior options, similarly to area II in
Figure~\ref{fig:stabilityN3sym}.
The bifurcation diagrams in the right panels show the effects of $r$
for fixed values of $\kappa$. When the best option has double quality
than the inferior options (i.e., $\kappa=0.5$, see the bottom-right
panel), a low value of $r$ guarantees selection of the best option,
whereas a sufficiently high $r$ may result in incorrect decisions by
selecting any of the inferior options (which are considerably worse
than the best one).
As the inferior options become comparable to the superior one, the
range of values of $r$ in which there exists a single stable
equilibrium in favour of the best options gets reduced (see the
middle-right panel for $\kappa=0.9$ in
Figure~\ref{fig:dynamicsN3asym}), up to the point that there is no
value of $r$ in which the choice of the superior option is the unique
solution (see the top-right panel for $\kappa=0.97$ in
Figure~\ref{fig:dynamicsN3asym}). In this case, however, there is
little difference in quality between the superior and inferior
options, and the system dynamics are similar to the symmetric case in
which it is most valuable to break a decision deadlock, hence to
choose a sufficiently high value of $r$.

\begin{figure}
  \centering 
  \subfigure[]{ \label{fig:stabilityRKmanyN}
    \includegraphics[width=0.43\linewidth, trim=0 0 0 0cm]{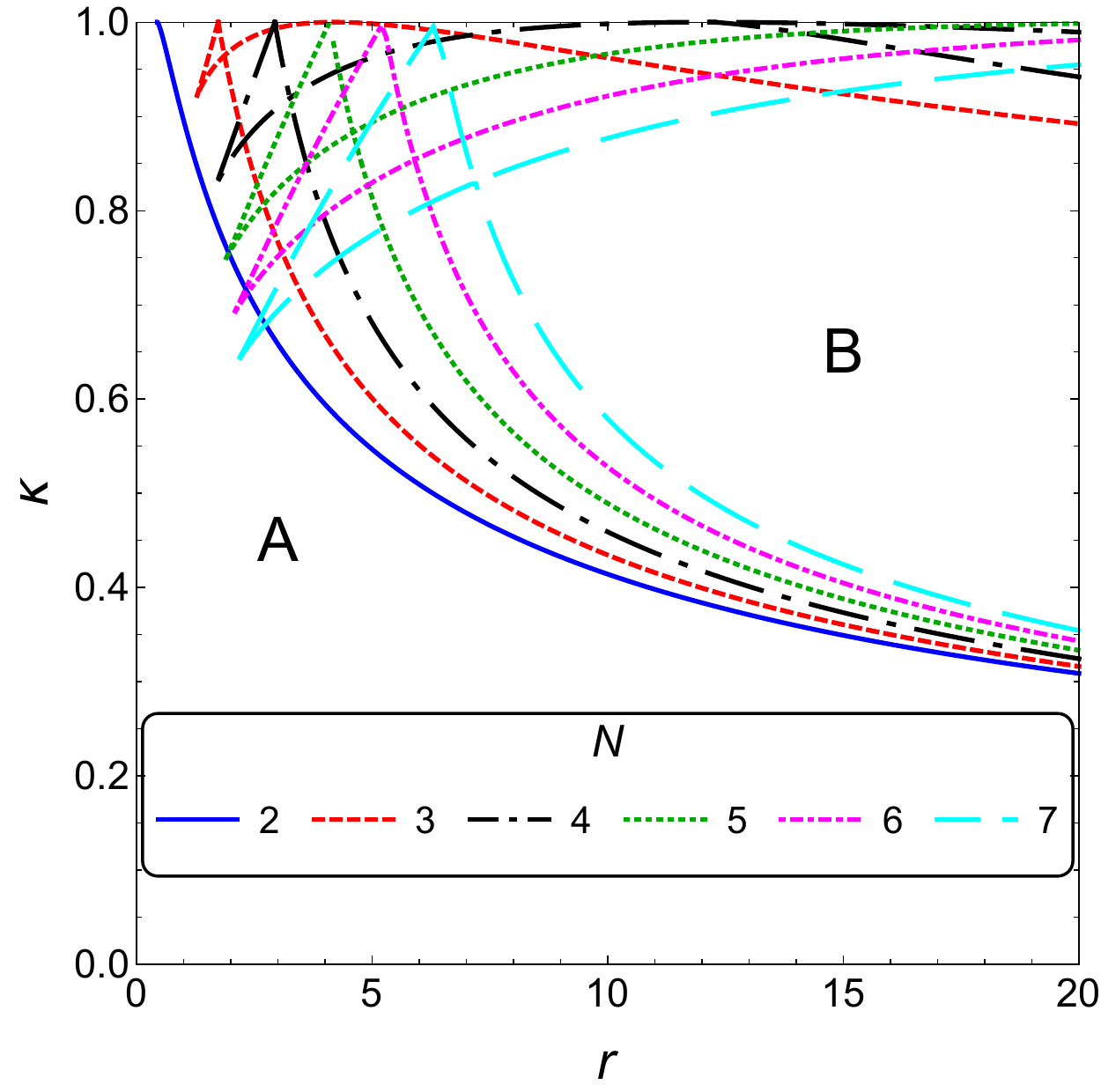}}
   \subfigure[]{ \label{fig:densityMapMaxKappa}
     \includegraphics[width=0.485\linewidth, trim=0 0 0 0cm]{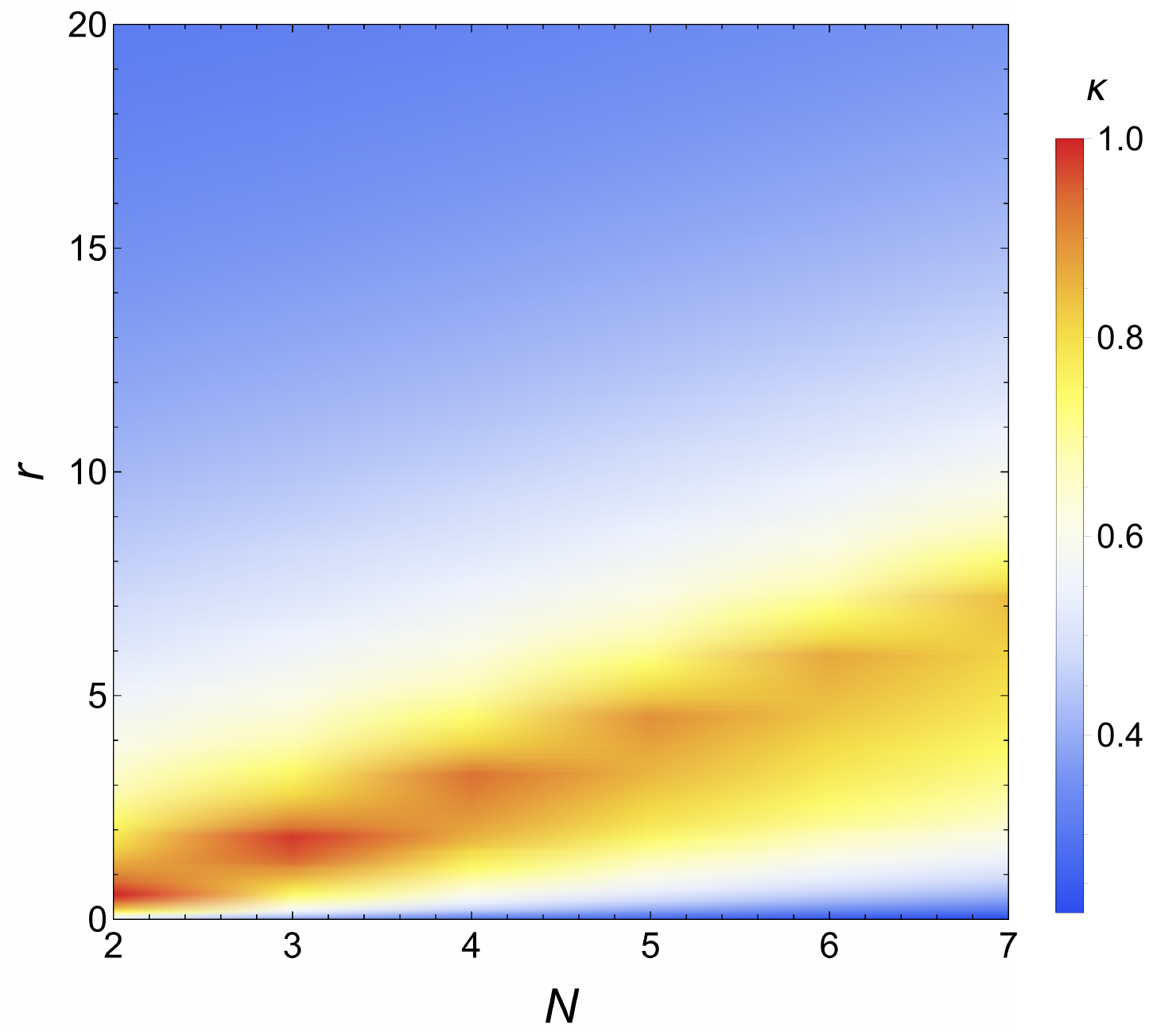}}
   \caption{\subref{fig:stabilityRKmanyN} Stability diagram for best
     option quality $v=5$ in the parameter space \mbox{$r,\, \kappa$}
     for varying number of options \mbox{$N \in \{2,\dots,7\}$}. For
     each option, the system has five possible phases that are
     consistent with the phases described in caption the of
     Figure~\ref{fig:dynamicsN3asym}. Here we label only areas A
     (monostability) and B (multistability) to facilitate
     readability. \subref{fig:densityMapMaxKappa} Maximum value of
     $\kappa$ as a function of $N \in \{2,\dots,7\}$ and
     $r \in (0,20]$ for which the system has a unique attractor for
     the selection of the best quality option, defined as the best
     option attracting commitment from at least 75\% of the total
     decision-making population.}
\label{fig:stabilityRKN234}
\end{figure}

The dynamics observed for $N=3$ options are consistent in the case of
$N>3$. 
Figure~\ref{fig:stabilityRKmanyN} shows the stability diagram for
varying number of options \mbox{$N \in \{2,\dots,7\}$} (see also
Figure~\ref{fig:compareStabDiag}). It is possible to note that areas D
and E get larger as $N$ increases, leading to a larger range of values
in which one or more stable decision states coexist with a stable
undecided state, up to the point that area C disappears for $N\geq
5$. This means that, as the number of inferior options increases, the
probability of making a wrong decision increases as well, especially
for high values of $\kappa$. To minimise the probability of wrong
decisions, the value of $r$ should be maintained as small as possible,
but still high enough to ensure that a decision is taken (i.e., with a
sufficiently large population committed to one option, see the density
map in Figure~\ref{fig:compareStabDiag}).
Finally, in Figure~\ref{fig:densityMapMaxKappa} we show how the
ability to solve hard decision problems varies with $r$ and $N$. To
this end, for each point in the space $r,N$, we show the highest value
of $\kappa$ for which there exists a unique attractor for the superior
option corresponding to at least 75\% of the population committed
(i.e., $x_1 \geq 0.75$). Figure~\ref{fig:densityMapMaxKappa}
demonstrates an approximately linear relationship between $r$ and $N$
for a given value of $\kappa$.

\section{Discussion}

We have analysed a model of consensus decision-making which exhibits
useful value-sensitive properties that have previously been described
for binary decisions \cite{Pais:pone:8:e73216:2013}, but generalises
these to decisions over three or more options. In order to preserve
these properties the single control parameter in the original model of
\cite{Pais:pone:8:e73216:2013}, the rate of cross-inhibition between
decision populations, is replaced by a parameter describing the
relative frequencies with which individual group members engage in
independent discovery and abandonment behaviours, compared to positive
and negative-feedback signalling behaviours. This new control
parameter is biologically meaningful and experimentally measurable, so
should be of interest for further empirical studies of house-hunting
honeybee swarms.

Previous work has investigated the role of signalling in collective
decision making in a somewhat different
framework. Galla~\cite{Galla:JTB:2010} has analysed a model of
house-hunting honeybees~\cite{List:PTB:2009} where the
cross-inhibition mechanism was not included. In this model, increasing
signalling (referred to as interdependence) allows the swarm to select
the best quality option more reliably. The interdependence parameter
modulates the strength of positive feedback; the higher the
interdependence is, the more a bee is influenced by other bees'
opinion in determining a change of commitment.
There are similarities and differences between the meaning of the
interdependence parameter and the signalling ratio $r$ that is
introduced in this paper. Similarly to \cite{Galla:JTB:2010,
  List:PTB:2009}, increasing the value of the ratio $r$ corresponds to
an increase in the signalling behaviour but, in contrast to previous
studies, $r$ is a weighting factor of both positive and the negative
feedback. However, note that positive and negative feedback are not
necessarily equal in our model, as these mechanisms are also modulated
by the option's quality. In agreement with \cite{Galla:JTB:2010,
  List:PTB:2009}, our results underline the importance of interactions
among honeybees in the nest-site selection process. However, given the
different meanings of the control parameters, we find that increased
signalling behaviour helps to break decision deadlocks (in case of
equal alternatives) but too high signalling might reduce the decision
accuracy when the decision has to be made among different quality
options.

We also note some similarities between our results and the bifurcation
analysis of a model of the collective decision making process in
foraging ants \emph{Lasius niger}~\cite{Nicolis:JTB:1999}.
This model describes the temporal evolution of the pheromone
concentration along $N$ alternative trails, each of which leads to a
different food source. The bifurcation parameter in the analysis is an
aggregate variable composed of the total population size, the options'
qualities and the pheromone evaporation rate. Not all of these
components are under the direct control of the decision maker, and
thus cannot be varied during the decision process.  In contrast, the
control parameter in our analysis, the signalling ratio $r$, can be
modulated in a decentralised way by the individual bees.
Comparing the bifurcation diagrams for deadlock breaking of Fig.~3(a)
in \cite{Nicolis:JTB:1999} with Fig.~\ref{fig:BifN4}, the
two models present similar dynamics.
The authors also present a hysteresis loop as a function of relative
food source quality (Fig.~4 in \cite{Nicolis:JTB:1999}), which is
similar to that found as a function of relative nest-site quality in
\cite{Pais:pone:8:e73216:2013} (Fig. 5). Collective foraging over
multiple food sources is a fundamentally different problem to
nest-site selection, with exploitation of multiple sources frequently
preferred in the former whereas convergence on a single option is
required in the latter
\cite{Marshall:JRSInterface:6:1065:2009}. Nevertheless it could be
interesting to make further comparisons of the dynamics of the model
presented here and other nonlinear dynamical models exhibiting
qualitatively similar behaviour.
  
A crucial point in our model is that honeybees need to interact at a
rate that is high enough to break decision deadlock in the case of
equal options, in addition to the influence of nest-site qualities.
This follows from our analysis of the symmetric case
(Section~\ref{sec:results-sym}), where we observed that high
signalling ratio $r$ allows the system to break the decision deadlock
and to select any of the equally best options. However, the analysis
of the asymmetric case (Section~\ref{sec:results-asym}) revealed that
a frequent signalling behaviour may have a negative effect on the
decision accuracy, and low $r$ values should be preferred to have a
systematic choice of the best available option.
These results suggest that a sensible strategy may be to increase $r$
through time.  An organism may start the decision process applying a
conservative strategy which reduces unnecessary costs of frequent
signalling behaviour and that, at the same time, allows quickly and
accurately to select the best option if it is uniquely the
best. Otherwise, in the case of a decision deadlock (due to multiple
options having similar qualities), the system may increase its
signalling behaviour in order to break symmetry and converge
towards the selection of the option with the highest quality. This
strategy is reminiscent of the suggested strategy of increasing
cross-inhibition over time to spontaneously break deadlocks in binary
decisions~\cite{Pais:pone:8:e73216:2013}.
Further theoretical evidence supporting such a strategy comes from the
bifurcation diagrams presented in the middle- and top-right panels in
Figure~\ref{fig:dynamicsN3asym}, corresponding to asymmetric case with
$N=3$ similar options, with $\kappa=0.9$ and $\kappa=0.97$,
respectively (see also Figure~\ref{fig:bifurcationsManyNsAsym} for
further bifurcation diagrams with $N \in \{4,5,6,7\}$). In these
cases, an incremental increase in $r$ would allow the system to
converge accurately towards the best option. In contrast, immediately
starting the decision process with a high value of $r$ might decrease
the decision accuracy. For instance, in
Figure~\ref{fig:dynamicsN3asym}~(right-center), starting with low
values of $r$ (i.e., $r<2.1$) would bring the system to the stable
attractor (blue line) with less than half of the population committed
to the best option. A gradual increase of $r$ lets the process follow
the (blue, stable) solution line which leads to the selection of
option $1$. On the other hand, a process that starts from a totally
uncommitted state with a value of $r>2.1$ may end in the basin of
attraction corresponding to selection of an inferior option, as a
consequence of stochasticity of the decision process.
Such a strategy could easily be implemented in a decentralised manner
by individual group members slowly increasing their propensity to
engage in signalling behaviours over time; such a direction of change,
from individual discovery to signalling behaviour, is also consistent
with the general requirement of a decision-maker to gather information
about available options, but then to begin restricting consideration
to these rather than investing time and resources in the discovery of
further alternatives.
Theorists and empiricists have previously concluded that honeybee
swarms achieve consensus through the \textit{expiration of
  dissent}~\cite{Seeley:BES:2003}, which occurs as bees apparently
exhibit a spontaneous linear decrease in number of waggle runs for a
nest over time~\cite{Seeley:BehavEcolSociobiol:45:19:1999}. However,
the discovery of stop-signalling in swarms requires that this
hypothesis be re-evaluated, since increasing contact with
stop-signalling bees over time will also decrease expected waggle
dance duration~\cite{Seeley:Science:2012}. Field observations report
that recruitment decreases over time in easy decision problems while
it increases overall in difficult problems (\textit{e.g.} five
equal-quality nests)~\cite{Seeley:AS:2006}. Further theoretical work
with our model would reveal whether it is capable of explaining these
empirically-observed patterns.

Our analyses also suggest an optimal stable signalling ratio that the
decision-making system might converge to. While the level of
signalling required to break deadlock between $N$ equal options
increases quadratically with $N$ (Figure~\ref{fig:bifOnNsym}), the
level of signalling that optimises the discriminatory ability of the
swarm in best-of-N scenarios increases only linearly
(Figure~\ref{fig:densityMapMaxKappa}). Optimising best-of-N decisions
therefore seems at odds with optimising equal alternatives
scenarios. However in natural environments the probability of
encountering $N$ (approximately) equal quality nest options will
decrease rapidly with $N$. On the other hand the best-of-N scenario
here, while still less than completely realistic, should still provide
a better approximation to the naturalistic decision problems typically
encountered by honeybee swarms. Our analysis shows that the level of
signalling that swarms converge to may be tuned appropriately by
evolution according to typical ecological conditions, namely the
number of potentially suitable nest sites that are typically available
within flight distance of the swarm. Swarms of the European honeybee
\textit{Apis mellifera} are able to solve the best-of-N problem with
one superior option and four inferior options
\cite{Seeley:BehavEcolSociobiol:2001}, presumably reflecting the
typical availability of potential nest sites in their ancestral
environment.

While our model is inspired by nest-site selection in honeybee swarms,
we feel its relevance is potentially much greater. For example, as
mentioned in the introduction, decision-making in microbial
populations may share similarities with decisions by social insect
groups \cite{RossGillespie:fmicb:5:54:1:2014}. In addition
cross-inhibitory signalling is a typical motif in intra-cellular
decisions over, for example, cell fate
\cite{Nene:pone:7:0032779:2012}, and single cells can exhibit decision
behaviour similar to Weber's law \cite{Ferrell:MolCell:36:724:2009,
  Goentoro:MolCell:36:894:2009}.  Weber's law describes how the
ability to perceive the difference between two stimuli varies with the
magnitude of those stimuli, and may have adaptive
benefits~\cite{Akre:TrEcolEvol:29:291:2014}. Several authors have also
noted similarities between collective decision-making and organisation
of neural decision circuits, where inhibitory connections between
evidence pathways are also typical \cite{Couzin:Nature:445:715:2007,
  Marshall:CurrBiol:19:R395:2009, Couzin:TrendsCognSci:13:36:2009,
  Passino:BehavEcolSociobiol:62:401:2007,Marshall:JRSInterface:6:1065:2009}.
Similarly, neural circuits following the winner-take-all principle
have dynamics regulated by the interplay of excitatory and inhibitory
signals and present interesting analogies to the present model
\cite{Douglas1995, Rutishauser2011}. Since organisms at all levels of
biological complexity must solve very similar statistical decision
problems that relate to fitness in very similar ways, we feel there is
definite merit in continuing to pursue the analogies between
collective decision-making models such as that presented here, and
models developed in molecular biology and in neuroscience. Finally, we
suggest that the simplicity of the model presented here and its
adaptive decision-making characteristics might inform the design of
artificial decentralised decision-making systems, particularly in
collective robotics
(\textit{e.g.}~\cite{Leonard:AnnuRevContr:38:171:2014,
  Reina:10:pone:0140950:2015, Reina:SwarmIntelligence:9:75:2015,
  Reina:DARS:2016}) and in cognitive radio networks
(\textit{e.g.}~\cite{Trianni:ICC:2016}).

\section{Acknowledgments}
This work was funded by the European Research Council (ERC) under the
European Union's Horizon 2020 research and innovation programme (grant
agreement number 647704). Vito Trianni acknowledges support by the
European Commission through the Marie Sk\l{}odowska-Curie Career
Integration Grant ``DICE, Distributed Cognition Engineering'' (Project
ID: 631297).

\section*{Appendices}

The appendixes are organised in five sections. In
Appendix~\ref{sec:models}, we present the complete model in all the
parameterisations discussed in the article (from the most general to
the most specific). Then, we report the reduced model in a similar set
of parameterisations. 
In Appendix~\ref{sec:pais-deadlock}, we show that the parameterisation
used in the literature~\cite{Pais:pone:8:e73216:2013} cannot break the
decision deadlock in the symmetric case when the number of options is
larger than two.
In Appendix~\ref{sec:binary}, we study the dynamics of the system in
the selected parameterisation for the binary case, i.e., $N=2$.
In Appendix~\ref{sec:bifurcations}, we report the formulas of the two
main bifurcation points for the symmetric case. This formula is
particularly significant because it is valid for any number of
options.  In Appendix~\ref{sec:dynamics}, we report additional results
on the system dynamics: we report additional analysis performed on the
system deciding between three options, and we show that the results
for $N=3$ options are qualitatively similar for $N>3$.

\appendix

\section{Complete model and reduced model}
\label{sec:models}

The general model for $N$ options is:
\begin{align} 
  \left\{ 
    \begin{aligned}
      \frac{dx_i}{dt} &= \gamma_i\, x_u - \alpha_i\, x_i + \rho_i\, x_u\, x_i 
                        - \sum_{j=1}^N\, x_j\, \beta_{ji}\, x_i \,,\qquad  i \in \{1,\,\dots\,,\,N\} \,, \\
      x_u &= 1 -  \sum_{i=1}^N x_i 
    \end{aligned}
  \right.
  \label{eq:fullSystem}
\end{align}
where $x_i$ represents the subpopulation committed to option $i$ and
$x_u$ the uncommitted subpopulation. $\gamma_i$ represents the
discovery rate for option $i$, $\alpha_i$ the abandonment rate for
option $i$, $\rho_i$ the recruitment rate for option $i$ and
$\beta_{ji}$ the cross-inhibition from subpopulation $j$ to
subpopulation $i$.

We introduce a first parameterisation as:
\begin{equation}
  \gamma_i = k \, v_i \qquad
  \alpha_i = k \, v_i^{-1} \qquad
  \rho_i = h \, v_i \qquad
  \beta_{ii}=0 \qquad
  \beta_{ij} = \beta
  \label{eq:half-param}
\end{equation}
with $i \neq j$. By applying Equation~\eqref{eq:half-param} in
\eqref{eq:fullSystem}, we obtain:
\begin{align} 
  \left\{ 
    \begin{aligned}
      \frac{dx_i}{d\tau} &= v_i\, x_u - \frac{x_i}{v_i} + r \, v_i \, x_u\, x_i 
                        - \sum_{j=1,\, j\ne i}^N\, x_i\, \beta \, x_j \,,\qquad  i \in \{1,\,\dots\,,\,N\} \,, \\
      x_u &= 1 -  \sum_{i=1}^N x_i 
    \end{aligned}
  \right.
  \label{eq:sys-half-param}
\end{align}
where $r=h/k$ is the ratio of interaction over spontaneous
transitions, and $\tau = k\,t$ is the dimensionless time.  The
parameterisation of Equation~\eqref{eq:half-param} is a generalisation
of the one proposed in the literature~\cite{Pais:pone:8:e73216:2013},
since, using $r=1$, the system~\eqref{eq:fullSystem} reduces to the
old one, and thus displays the same dynamics.

\begin{figure}[t]
  \centering
    \includegraphics[width=0.666\linewidth]{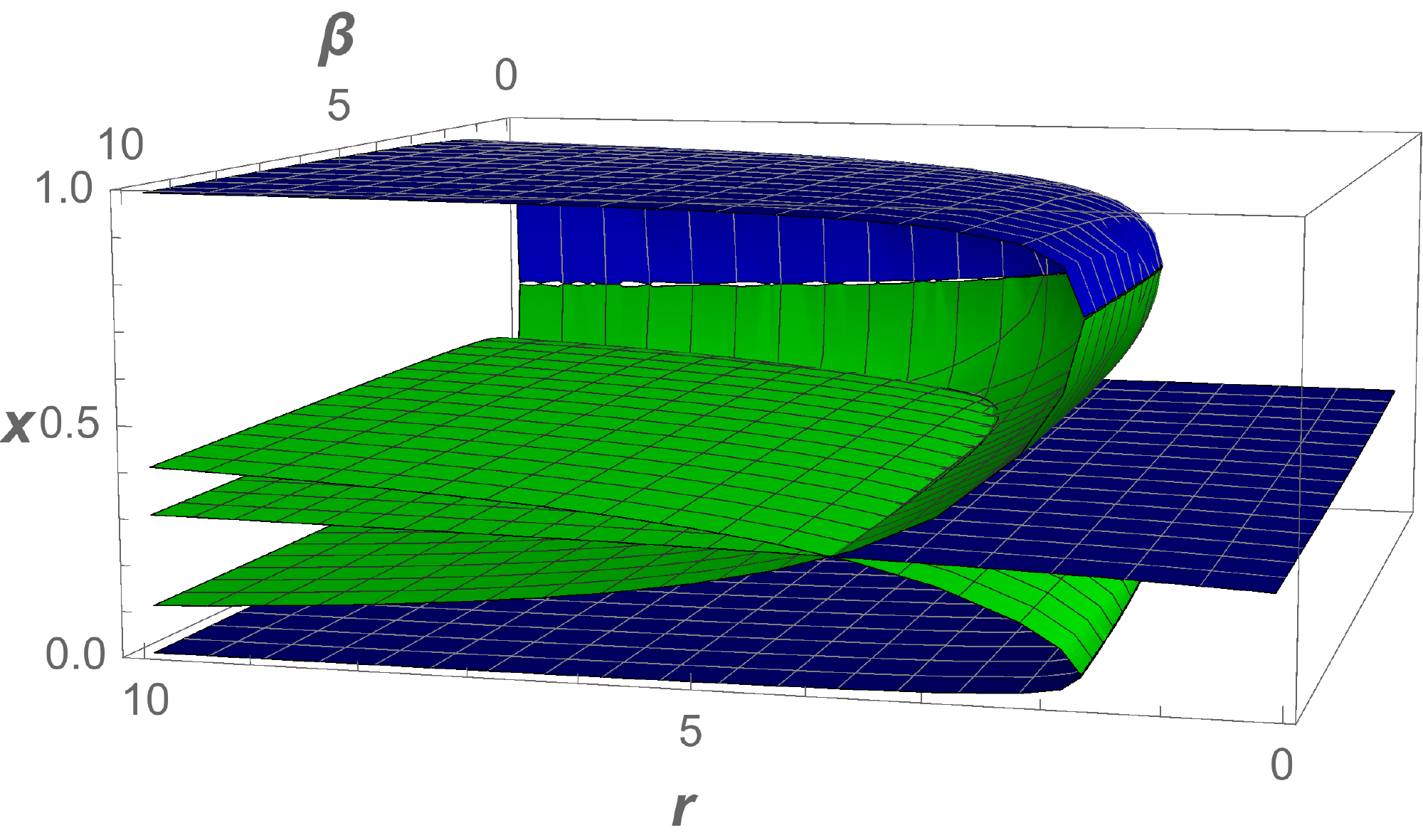}
    \caption{Bifurcation diagram in 3D of the
      system~\eqref{eq:sys-half-param} with $N=3$ equal-quality
      options (i.e., $v_1=v_2=v_3=v$) as a function of
      $r=h/k \in (0,10]$ and $\beta \in (0,10]$.  The vertical axis
      shows $x \in [0,1]$, which represents the proportion of bees
      committed to one of the three identical options. Blue surfaces
      represent stable equilibria, while green surfaces are unstable
      equilibria. We can see that for $r=1$, the decision deadlock is
      stable for any tested values of $\beta$. See
      Section~\ref{sec:pais-deadlock} for a formal proof of the
      decision deadlock for $r=1$ and $N=3$.}
\label{fig:Bifurcation3DforN2}
\end{figure}

This intermediate steps allows us to visualise that for $r\le1$ there
is no value of $\beta$ that allows to break the decision deadlock in
the case of $N>2$ same-quality options (see
Figure~\ref{fig:Bifurcation3DforN2}). This result motivates the change
of parameterisation with respect to previous
work~\cite{Pais:pone:8:e73216:2013}.  Additional analyses that confirm
the presence of the decision deadlock for values of $r=1$ are provided
in Appendix~\ref{sec:pais-deadlock}.

We modify the parameterisation of Equation~\eqref{eq:half-param} by
linking the signalling behaviours (recruitment and cross-inhibition)
with the same value. The modified parameterisation is:
\begin{equation}
  \gamma_i = k \, v_i \qquad
  \alpha_i = k \, v_i^{-1} \qquad
  \rho_i = h \, v_i \qquad
  \beta_{ij} = h \, v_i 
  \label{eq:params-supp}
\end{equation}
and by applying Equation~\eqref{eq:params-supp} in~\eqref{eq:fullSystem},
we obtain:

\begin{align}
  \left\{ 
  \begin{aligned}
    \frac{dx_i}{d\tau} &= v_i\, x_u - \frac{x_i}{v_i} +
    r\,v_i\, x_i \left[x_u - \sum_{j\neq i}\,\kappa_{ji}\, x_j \right] \,,\qquad  i,\,j = 1,\,...\,,\,N \,, \\
    x_u &= 1 - \sum_{i=1}^N x_i
  \end{aligned}
  \right.
  \label{eq:sys-new-param}
\end{align}
where $\kappa_{ij}=v_i/v_j$ the ratio between options's values (and
$\tau = k\,t$, again, is the dimensionless time).

\vspace{0.7cm}
\textbf{The reduced model.}
In this study, we investigate the scenario in which there is one
superior option and $N-1$ equal-quality inferior options.
Assuming that the best option is the option $1$, the
Equation~\eqref{eq:fullSystem} can be simplified through the following
variable change:
\begin{equation}
    \label{eq:var-change-general}
    x_{A}=x_{1}\qquad x_B=\sum_{i=2}^Nx_i, \qquad 
    \lambda_1 = \lambda_A \qquad \lambda_i = \lambda_B
    \qquad \lambda \in \{\gamma,\alpha,\rho,\beta\} 
  \qquad i \in \{2,\dots,N\} \,.
\end{equation}
By applying Equation~\eqref{eq:var-change-general} to the complete
system~\eqref{eq:fullSystem}, we obtain:
\begin{align}
  \left\{ 
\begin{aligned}
  \frac{dx_A}{dt} &= \gamma_A\, x_u - \alpha_A \, x_A +
  \rho_A \, x_A \, x_u - \beta_B \, x_A \, x_B  \,, \\
  \frac{dx_B}{dt} &= (N-1)\,\gamma_B \, x_u - \alpha_B \, x_B +
  \rho_B \, x_B \, x_u - \frac{N-2}{N-1}\,\beta_B \, x_B^2 - x_A \, x_B
  \beta_A \,, \\
  x_u &= 1 - x_A - x_B \,,
\end{aligned}
 \right.
\label{eq:reducedSys-gen}
\end{align}  

Similarly, Equation~\eqref{eq:sys-new-param} can be simplified through
the following variable change:
\begin{equation}
    \label{eq:var-change}
    x_{A}=x_{1}\qquad x_B=\sum_{i=2}^Nx_i, \qquad v=v_1, \qquad
    \kappa=\frac{v_1}{v_{i}} \qquad
    v_i=\kappa \, v, \qquad i \in \{2,\dots,N\} \,.
\end{equation}
By applying Equation~\eqref{eq:var-change} to the complete
system~\eqref{eq:sys-new-param}, we obtain:
\begin{align}
  \left\{ 
\begin{aligned}
\frac{dx_A}{d\tau} &= v\, x_u - \frac{x_A}{v} + 
r\,v\, x_A \left[x_u - \kappa\, x_B \right] \,, \\
\frac{dx_B}{d\tau} &= (N-1)\,\kappa\,v\, x_u - \frac{x_B}{\kappa\,v} + 
r\,v\, x_B \left[\kappa\,\left(x_u - \frac{N-2}{N-1}\,x_B\right) - x_A \right] \,, \\
x_u &= 1 -  x_A - x_B \,,
\end{aligned}
 \right.
\label{eq:reducedSys-simple}
\end{align}  

\section{Need for a novel parameterisation: Decision deadlock for
  $N=3$}
\label{sec:pais-deadlock}

In this appendix, we show that the model of
Equation~\eqref{eq:sys-half-param} with $r=1$ and $N=3$ cannot break
the decision deadlock for any values of $\beta \ge 0$.

To prove this, we start from the reduced system given in
Equation~\eqref{eq:reducedSys-gen} (we could also use the full
three-dimensional system but due to the higher number of equilibria
this is more difficult).  Note that Equation~\eqref{eq:reducedSys-gen}
describes the reduced system before value-sensitivity is
introduced. In this form it is also equivalent to the case $r=1$.

We assume that $\alpha_A = \alpha_B =\alpha$,
$\beta_A = \beta_B =\beta$, $\gamma_A = \gamma_B =\gamma$, and
$\rho_A = \rho_B =\rho$.  If we calculate the equilibria we find that
there are up to four different points. One is always negative and
unstable.  Depending on the other three stationary states (the
symmetric solution, and two more) and their stability, we determine if
the decision maker ends up in decision-deadlock, or not.

Investigating the existence of the equilibrium points we can write down a generalised condition determining 
the existence of the two non-symmetric equilibrium solutions that evolve at the bifurcation point  
(cf.~\cite{Seeley:Science:2012, Pais:pone:8:e73216:2013}). This reads:
\begin{align}
\begin{aligned}
(-\alpha \beta + &2 \beta \gamma + \alpha \beta N - 3 \beta \gamma N + 
   \beta \gamma N^2 + \beta \rho - \beta N \rho)^2 \\
   &- 4 (\alpha \gamma - 2 \alpha \gamma N + \alpha \gamma N^2) (-2 \beta^2 + 
    \beta^2 N - \beta \rho + \beta N \rho) = 0 \,\,.
  \end{aligned}    
  \label{EqProofCond}
\end{align} 

We may resolve this equation with respect to $\beta$.\\

(1) If we let $N=2$ we obtain
\begin{align}
\begin{aligned}
\beta = \frac{4 \alpha \gamma \rho}{(\rho-\alpha)^2}\,\,,
  \end{aligned}
  \label{EqProofSeeley}    
\end{align} 
as in the original model in \cite{Seeley:Science:2012}. \\

(2) If we now introduce value-sensitivity, i.e. $v_1=v_2=v$ ($2$ equal options),
and let $N=2$, $\rho = v$, $\gamma = v$, $\alpha = 1/v$ we get:
\begin{align}
\begin{aligned}
 \beta = \frac{4 v^3}{(1-v^2)^2}\,\,,
  \end{aligned}
  \label{EqProofPais}    
\end{align} 
which coincides with the result reported in \cite{Pais:pone:8:e73216:2013}.\\

(3) If we let $N=3$ (and accordingly $v_1=v_2=v_3=v$ ($3$ equal options)), 
$\rho = v$, $\gamma = v$, $\alpha = 1/v$,
which is the extension from $2$ options (see model in \cite{Pais:pone:8:e73216:2013}) 
to $3$ options we obtain for $v > 1/2$:
\begin{align}
\begin{aligned}
 \frac{8 v^3}{1-4 v^2} < \beta < 0\,\,.
  \end{aligned}
  \label{EqProofOurWork}    
\end{align} 

In Eqs.~\eqref{EqProofSeeley} - \eqref{EqProofOurWork} we gave the
condition for the existence of the two stationary points which might
be reached by the decision-maker in addition to the symmetric
solution.  These are related to pitchfork ($N=2$) or limit point
($N=3$) bifurcations.  If the parameter $\beta$ does not range in
these intervals only the symmetric equilibrium is real and positive,
which is the condition for biological meaningful states.  This
symmetric equilibrium is also stable.  In particular,
Eq.~\eqref{EqProofOurWork} shows that $\beta$ needs to be negative to
make the stationary states in question occur.  As, on the other hand,
$\beta$ needs to be positive in order to describe cross-inhibition,
this case has to be excluded and hence we have shown that the
parametrisation introduced in \cite{Pais:pone:8:e73216:2013} cannot
describe decision-deadlock breaking for $3$ options, as only one
stable equilibrium exists (the symmetric solution)
for $r=1$ and all $\beta \ge 0$. \\

Also note that the quality values associated with the available options should be $v \ge 1$. Otherwise, some of
the available states may take negative values, which is not a biologically relevant solution. This applies to all 
the parametrisations mentioned above.

\section{Effects of the novel parameterisation for $N=2$}
\label{sec:binary}

\begin{figure}[t]
  \centering
  \subfigure[]{
    \label{fig:N2symmetric}
    \includegraphics[width=0.48\linewidth]{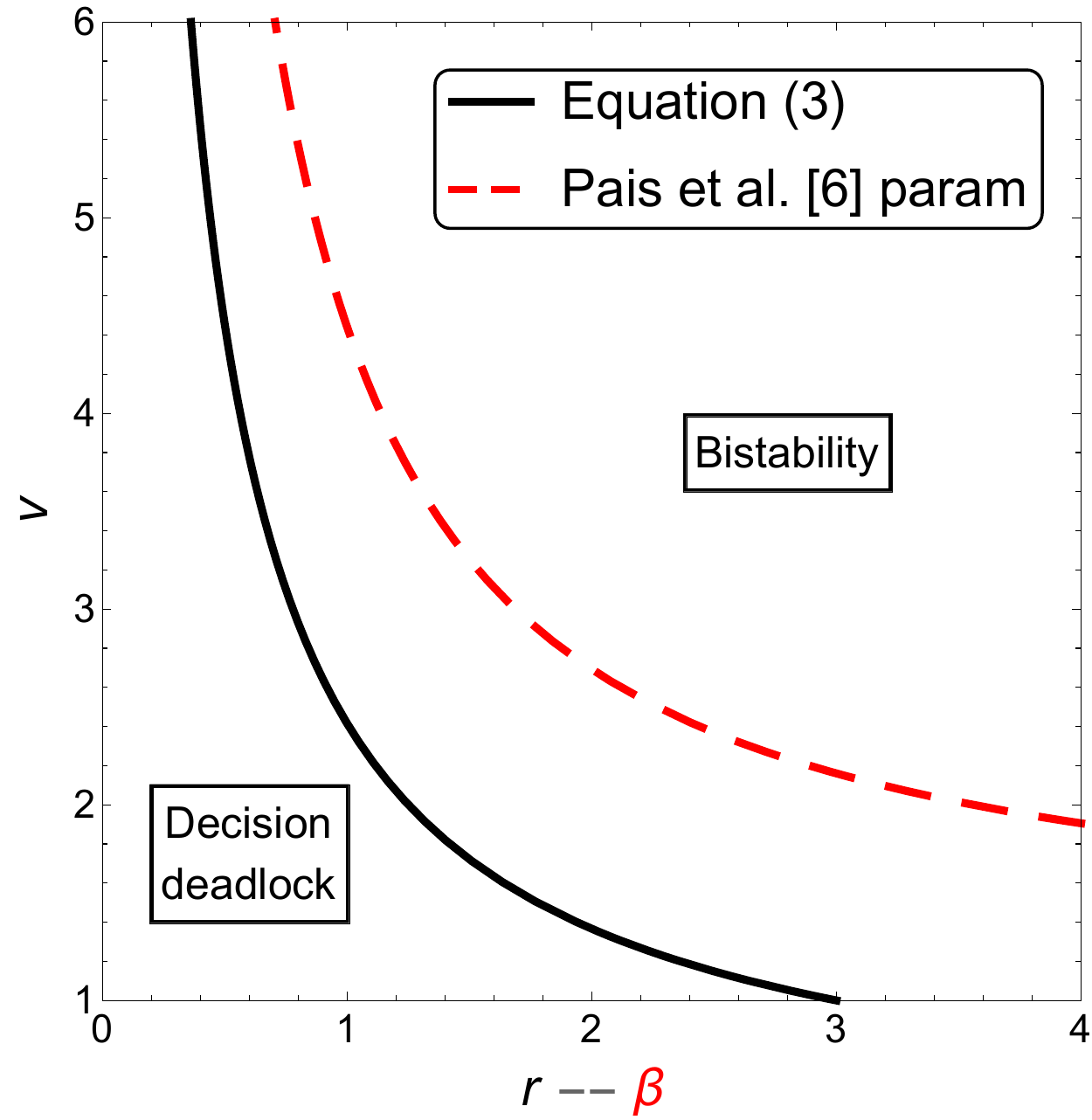}}
  \subfigure[]{\label{fig:weberN2}
    \includegraphics[width=0.48\linewidth]{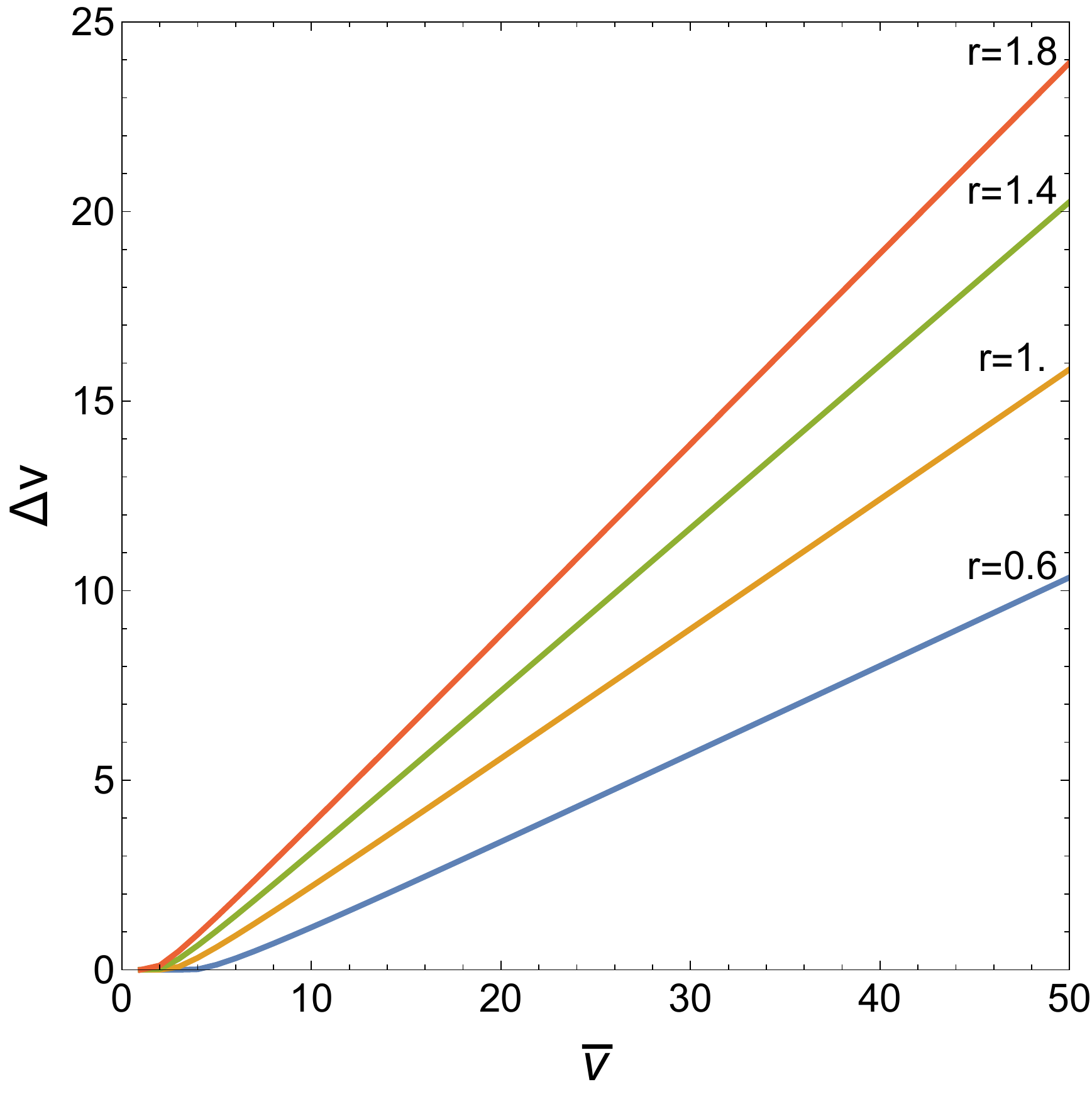}}
  \caption{\subref{fig:N2symmetric}~Comparison of the stability
    diagrams in the binary and symmetric case (i.e., $N=2$ and
    $v_1=v_2=v$) of the newly proposed parameterisation
    (Eq.~\eqref{eq:fullSysBestOfN}) and the previous
    work~\cite{Pais:pone:8:e73216:2013}. The bifurcation line that
    determines the two system phases is qualitatively similar, but the
    bifurcation parameter is different: In the previous work it is the
    cross-inhibition signal $\beta$, here it is the interaction ratio
    $r$. \subref{fig:weberN2}~Stability diagram of the
    system~\eqref{eq:fullSysBestOfN} as a function of the average
    quality $\bar{v}=(v_1+v_2)/2$ and the quality difference
    $\Delta v = |v_1-v_2|$ for varying $r \in \{0.6,1,1.4,1.8\}$, in
    the binary decision case. The lines show the relationship between
    the minimum quality difference to have the system with an unique
    attractor for the best option and the quality mean. This
    relationship resembles the Weber's law observed in psychological
    studies, with $r$ determining the coefficient. The results are
    similar to the ones obtained in~\cite{Pais:pone:8:e73216:2013},
    but using a different coefficient (in the previous work the
    coefficient was the cross-inhibition, $\beta$).}
\label{fig:comparisonNewParam}
\end{figure}

We study the dynamics of the systems~\eqref{eq:fullSysBestOfN} that
uses a novel parameterisation with respect to previous
work~\cite{Seeley:Science:2012, Pais:pone:8:e73216:2013}. We test if,
in the binary decision case (i.e., $N=2$), the system dynamics are
comparable to the dynamics reported in the literature.

Figure~\ref{fig:N2symmetric} shows a comparison of the stability
diagrams for the symmetric case of two options with equal value
$v$. The system dynamics are qualitatively similar but the bifurcation
parameter is different. In Pais et al., the bifurcation is determined
by the cross-inhibition $\beta$, while in our parameterisation it
is determined by the ratio of interaction/spontaneous transitions
$r = h/k$.

Additionally, Pais et al.~\cite{Pais:pone:8:e73216:2013} showed that
the cross-inhibition determines the minimum difference necessary to
discriminate between two similar quality options in a manner similar
to the Weber's law. We obtain similar results but using a different
parameter. In Figure~\ref{fig:weberN2} we show that the interaction
ratio $r$ determines the just noticeable difference.

\section{Bifurcations in the symmetric case}
\label{sec:bifurcations}

In case of $N$ equal-quality options, hereafter called the
\textit{symmetric} case, the values of every transition rate are the
same for both equation $A$ and $B$, i.e., $\gamma_A=\gamma_B=\gamma$,
$\alpha_A=\alpha_B=\alpha$, $\rho_A=\rho_B=\rho$ and
$\beta_A=\beta_B=\beta$.  The reduced system of
Equation~\eqref{eq:reducedSys-gen} becomes:

\begin{equation}
  \left\{ 
    \begin{array}{ll}
      \dot x_A&=\gamma x_U-\alpha x_A+\rho x_Ux_A-\beta x_Ax_B\\
      \dot x_B&=(N-1)\gamma x_U-\alpha x_B+\rho_Bx_Ux_B-\beta x_B(x_A +
\frac{N-1}{N-2}x_B)\\
      x_U &= 1 - x_A - x_B\\
    \end{array}\right.,
  \label{eq:approx-symmetric}
\end{equation}
System~\eqref{eq:approx-symmetric} undergoes two bifurcations. The
simplicity of Equation~\eqref{eq:approx-symmetric} allows us to
analytically derive the formula of the two bifurcation points:

\begin{align}
  \begin{aligned}
    \rho_1 &= \frac{\alpha  (2 \gamma  (N -1)+\sigma )+2 \sqrt{\alpha } \sqrt{\gamma }
   \sqrt{\alpha  (N -1)+\sigma  (N -2)} \sqrt{\gamma  (N -1)+\sigma
   }+\gamma  \sigma  (N -2)}{\sigma }\\
    \rho_2 &= \frac{\alpha  \left(\sqrt{\gamma } N  \sqrt{\gamma  N^2+4 \sigma }+\gamma 
   N^2+2 \sigma \right)+\sqrt{\gamma } \sigma  (N -2) \left(\sqrt{\gamma 
   N^2+4 \sigma }+\sqrt{\gamma } N \right)}{2 \sigma } \quad .
    \label{eq:bifurcation-general}
  \end{aligned}
\end{align} 

In the symmetric case, the system \eqref{eq:fullSysBestOfN} becomes:
\begin{align}
  \left\{ 
\begin{aligned}
\frac{dx_A}{d\tau} &= v\, x_u - \frac{x_A}{v} + 
r\,v\, x_A \left[x_u - x_B \right] \,, \\
\frac{dx_B}{d\tau} &= (N-1)\, v \, x_u - \frac{x_B}{v} + 
r\,v\, x_B \left[ x_u - \frac{N-2}{N-1}\,x_B - x_A \right] \,, \\
x_u &= 1 -  x_A - x_B \,,
\end{aligned}
 \right.
\label{eq:reducedSys-simple-sym}
\end{align} 
and undergoes two bifurcations at:
\begin{align}
  \begin{aligned}
    r_1 &= \frac{1}{v^2}-2+N+\frac{2\sqrt{2N-3}}{v}\\
    r_2 &= (N-3)N + 2 + \frac{1}{v^{2}} +
    \frac{N-1}{v}\sqrt{ (4 + v^2 (N - 2)^2)} \quad .
    \label{eq:bifurcations}
  \end{aligned}
\end{align}
Note, that here the bifurcation points are expressed as a function of
$N$, $r$ and $v$.


\section{System dynamics}
\label{sec:dynamics}

\textbf{Best of three.}
%

\begin{figure}[t]
  \centering
    \includegraphics[width=0.5\linewidth]{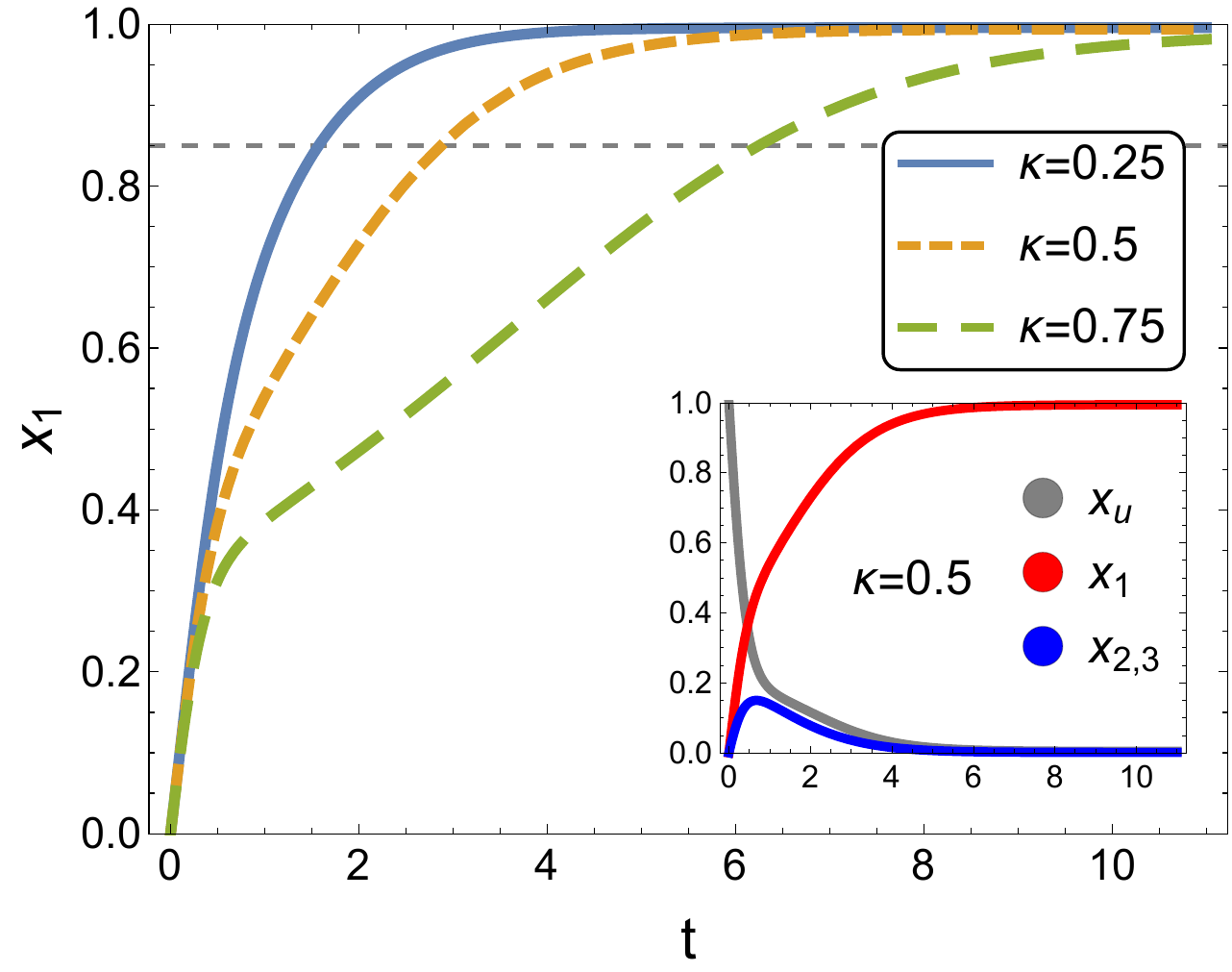}
    \caption{Time dependent solutions of the system of
      Equations~\eqref{EqSysOrigParam}-\eqref{eq:params} for $N=3$
      options, spontaneous transitions strength $k=0.1$, interaction
      transitions strength $h=0.3$, best option quality $v=10$, and
      varying inferior alternatives' quality as
      $\kappa \in \{0.25, 0.5, 0.75\}$. The main plot displays the
      dynamics of the population committed to the best quality option
      $x_1$; the inset shows the dynamics of all populations for
      $\kappa=0.5$, note that the populations committed for the
      inferior alternatives, $x_2$ and $x_3$, have overlaying
      trajectories. The horizontal dashed line shows an example quorum
      threshold \cite{Seeley:BehavEcolSociobiol:56:594:2004}.}
\label{fig:trajectoriesX1}
\end{figure}

Figure \ref{fig:trajectoriesX1} shows the time dependent solutions of
the system with $N=3$ options for varying values of
$\kappa \in \{0.25, 0.5, 0.75\}$. The plot shows the dynamics of the
population committed to the best quality option $x_1$. For decreasing
values of $\kappa$ the system converges faster to the stable
equilibrium $x_1=1$. The system parameters are in a plausible range
for the honeybee nest-site selection process leading to convergence
times that are comparable to field experiments, interpreting $t$ in
hour units \cite{Seeley:BehavEcolSociobiol:2001}.

\begin{figure}
\centering
  \subfigure{\label{fig:stabilityDiagramRK}
    \includegraphics[width=0.375\linewidth]{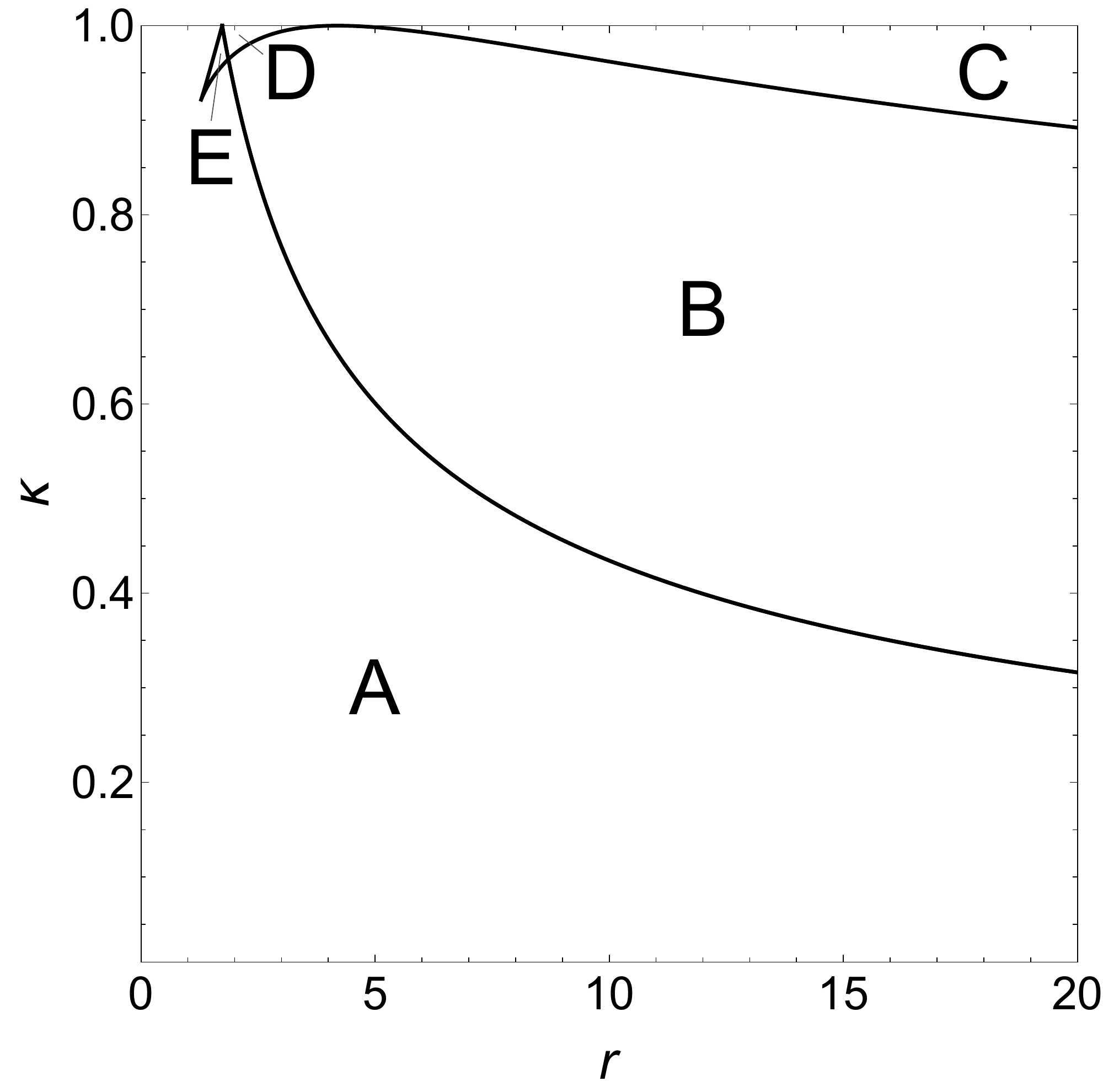}}
  \hspace{0.025\linewidth}
    \subfigure{\label{fig:phasePortraitA}
    \includegraphics[width=0.28\linewidth]{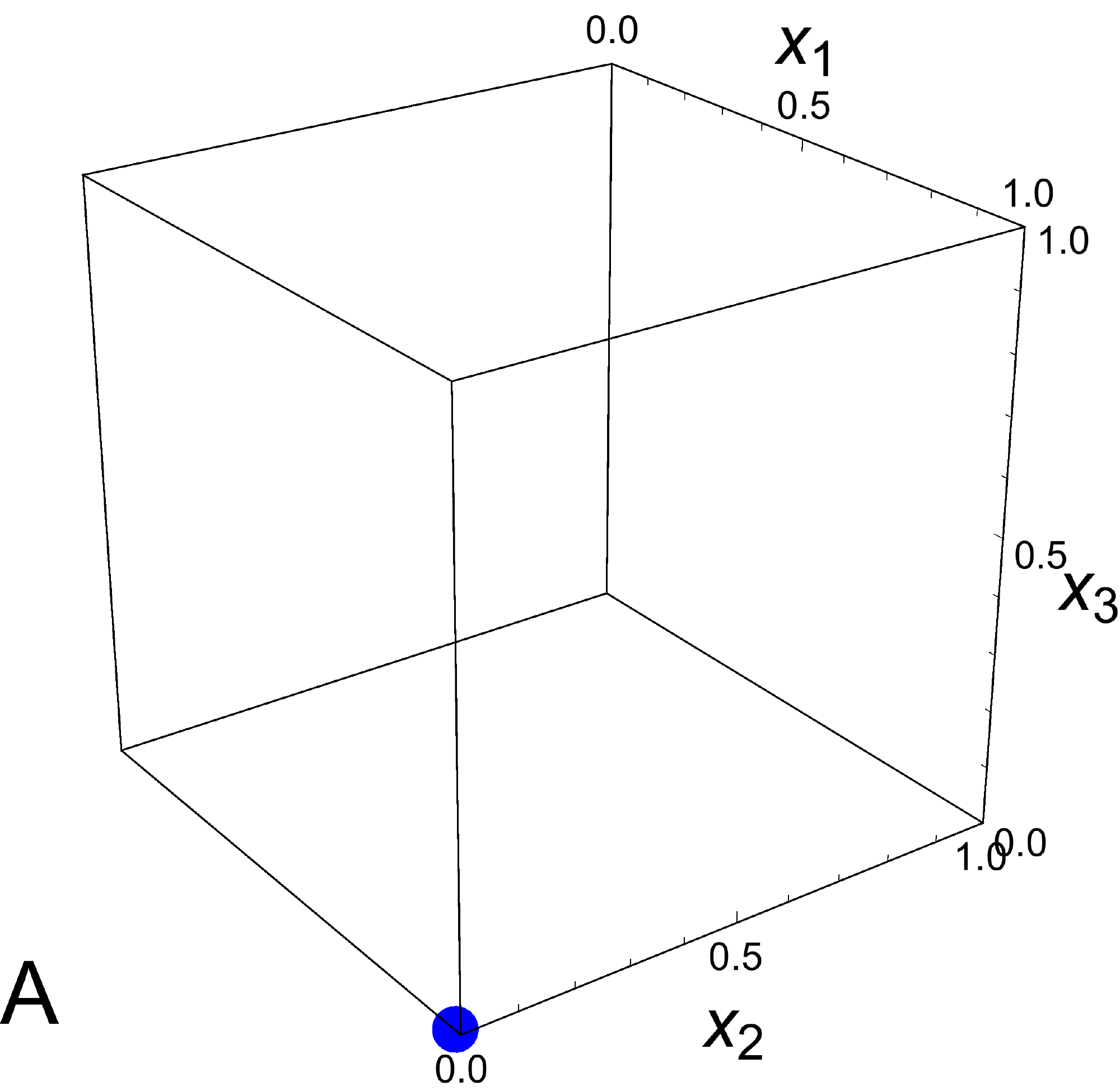}}
  \subfigure{\label{fig:phasePortraitB}
  \includegraphics[width=0.28\linewidth]{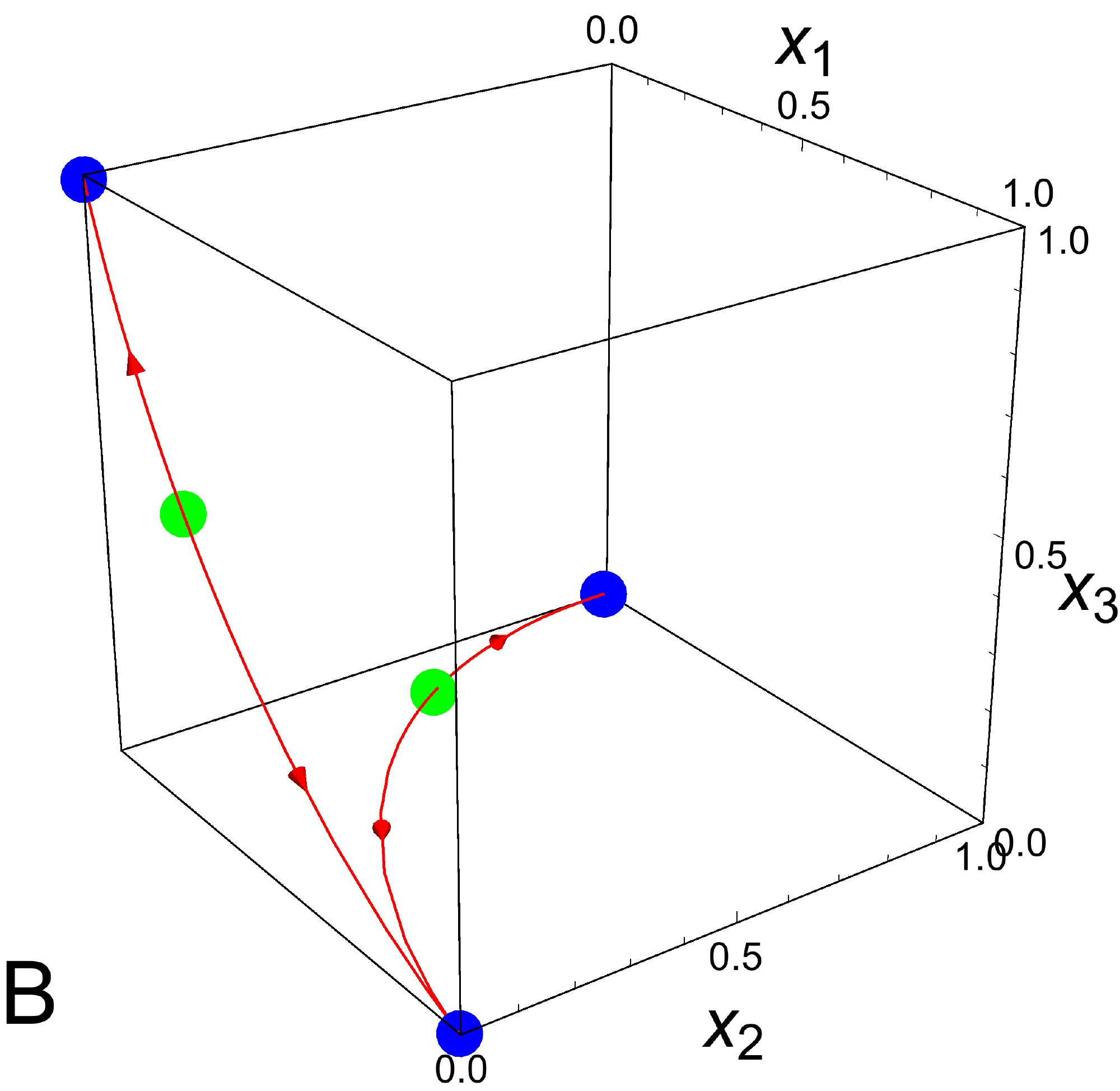}}\\

    \subfigure{\label{fig:phasePortraitC}
    \includegraphics[width=0.28\linewidth]{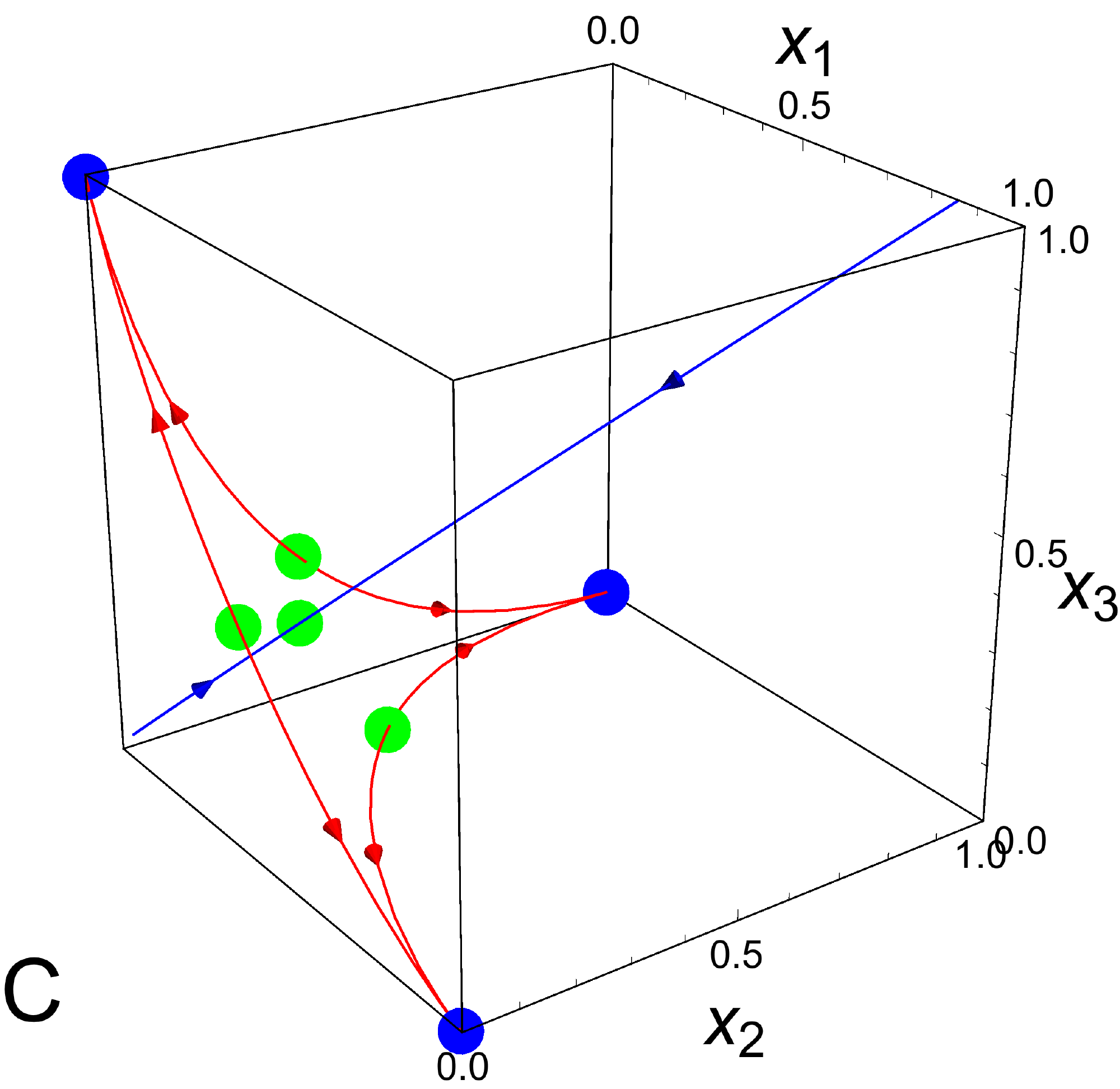}}
    \subfigure{\label{fig:phasePortraitD}
    \includegraphics[width=0.28\linewidth]{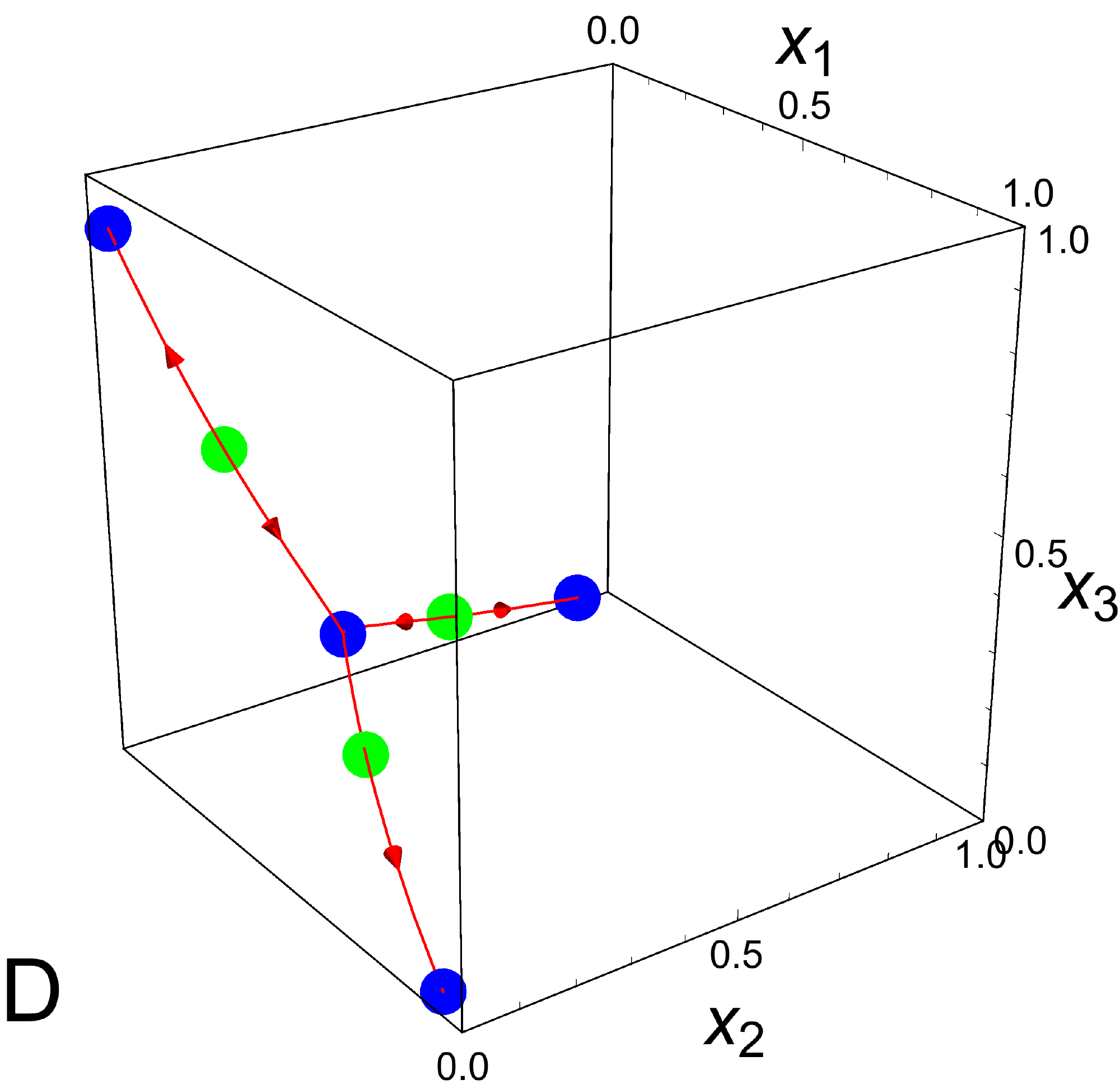}}
  \subfigure{\label{fig:phasePortraitE}
  \includegraphics[width=0.28\linewidth]{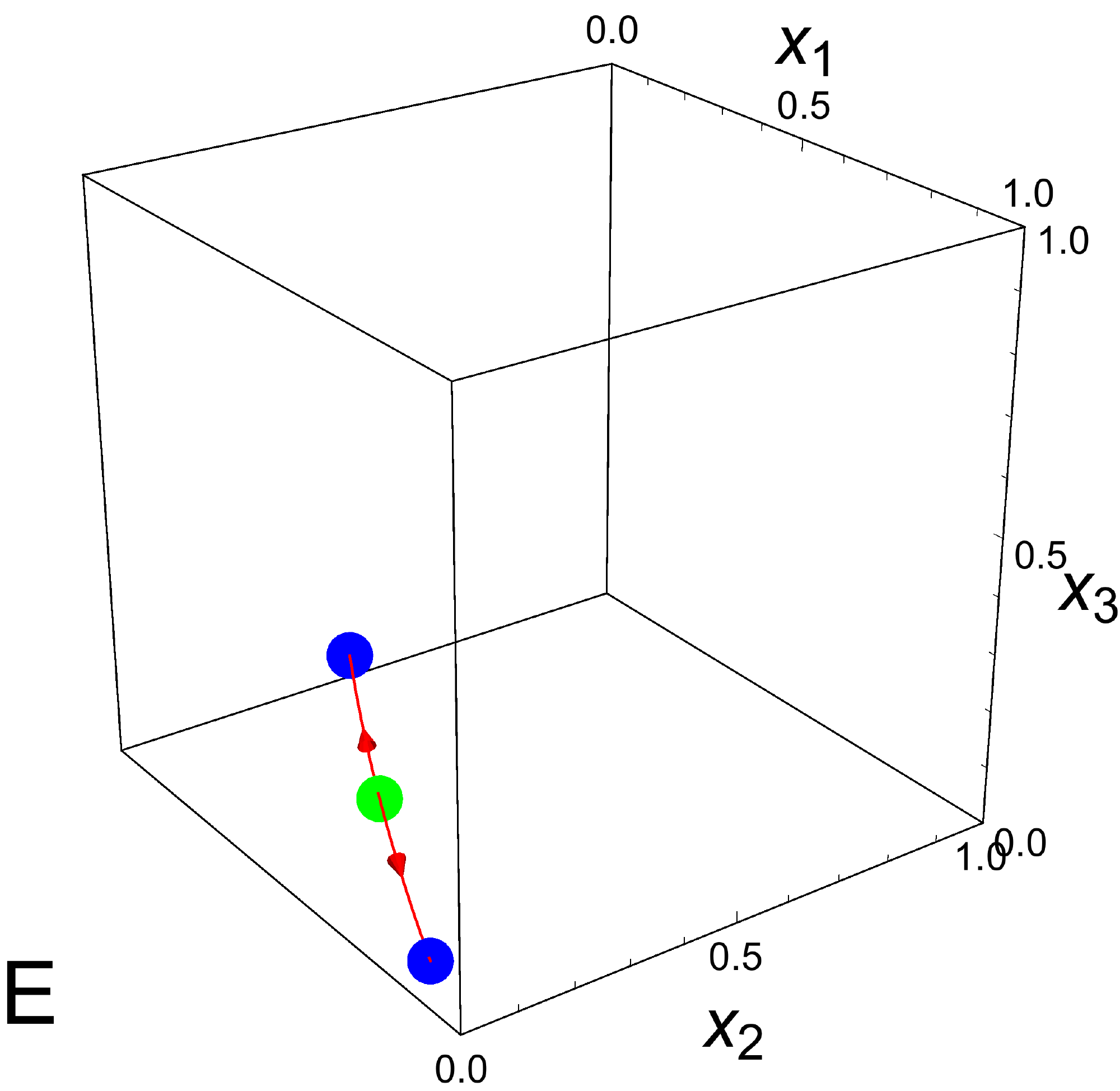}}
\caption{Dynamics of the system~\eqref{eq:fullSysBestOfN} in the case
  of $N=3$ options. In the top-left panel, we report the stability
  diagram in the parameter space $r$, $\kappa$. The plot shows
  that there are five possible system phases, labelled with letter from
  A to E. The other panels show a representative 3D phase portrait for
  each phase. The letter in the bottom-right of each phase portrait
  indicates which phase they represent.}
\label{fig:stabilityRK}
\end{figure}

In Figure~\ref{fig:dynamicsN3asym}, we identify five system phases
(labelled as \textbf{A}, \textbf{B}, \textbf{C}, \textbf{D} and
\textbf{E}) for the asymmetric case and $N=3$. In
Figure~\ref{fig:stabilityRK}, we report a representant 3D phase
portrait of the system~\eqref{eq:fullSysBestOfN} for each of the five
system phases.

\begin{figure}[t!]
  \centering
  \subfigure[]
  {\label{fig:stabDiagN4}
    \includegraphics[width=0.48\linewidth]{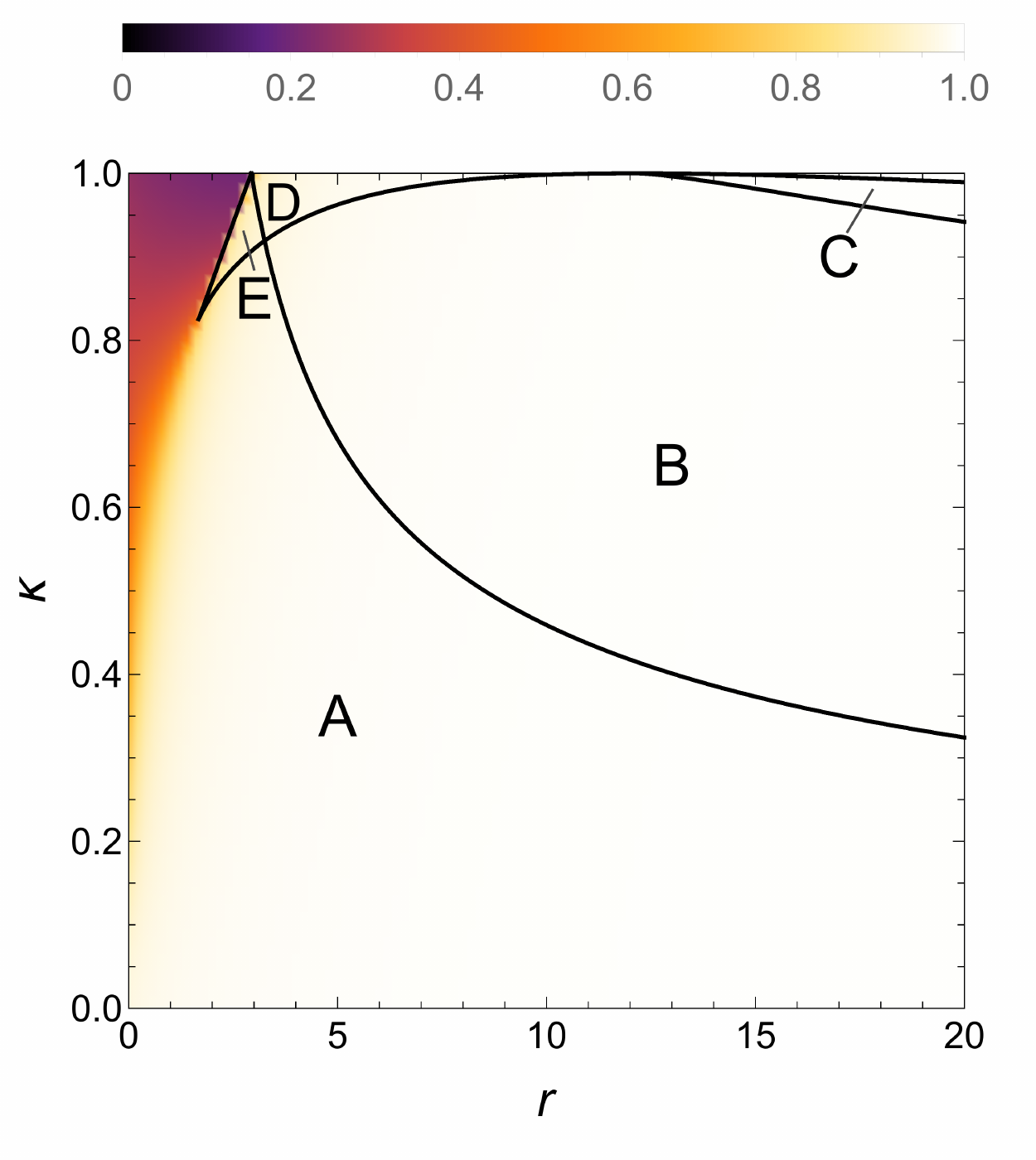}}
  \subfigure[]
  {\label{fig:stabDiagN5}
    \includegraphics[width=0.48\linewidth]{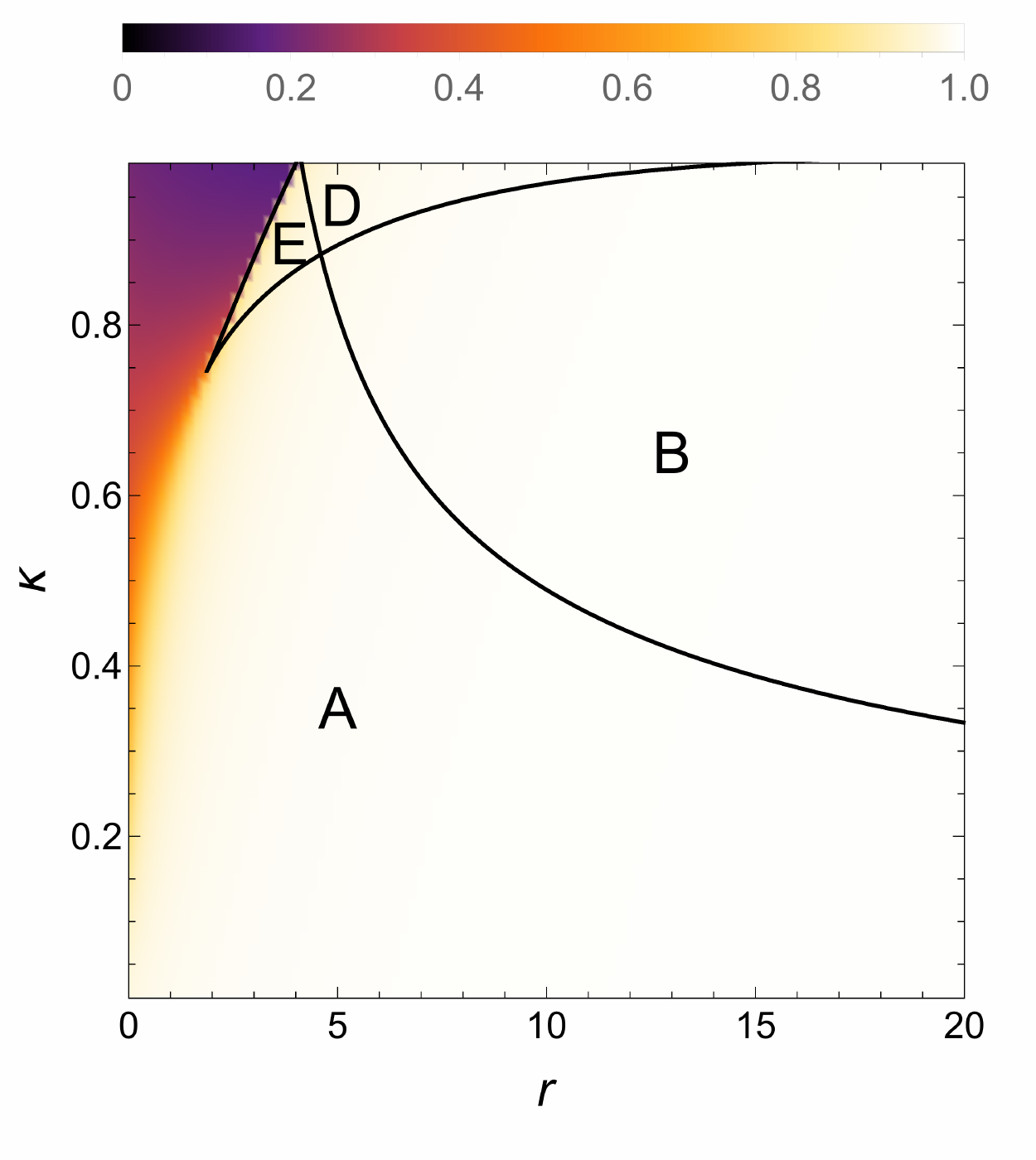}}\\
  \subfigure[]
  {\label{fig:stabDiagN6}
    \includegraphics[width=0.48\linewidth]{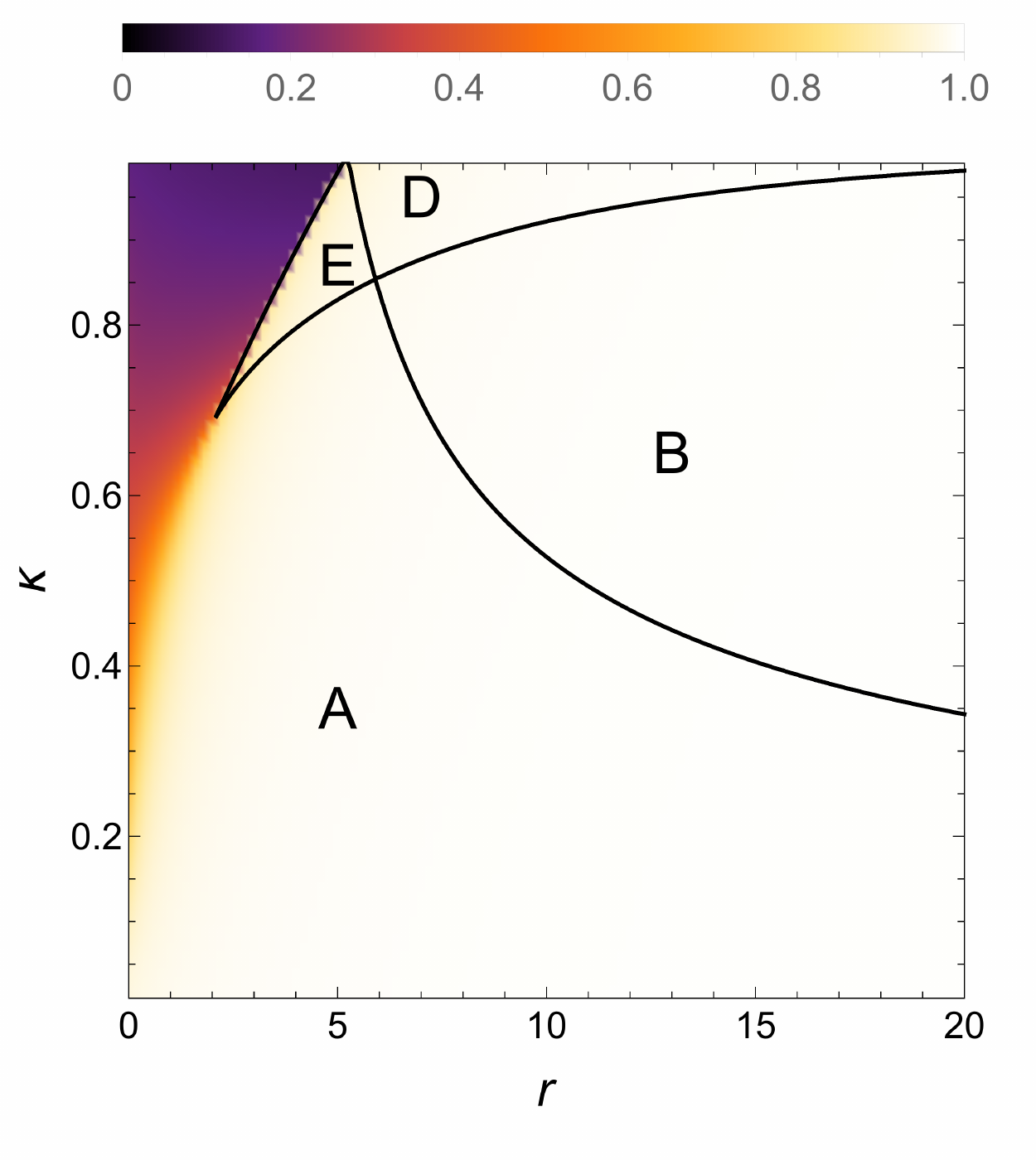}}
  \subfigure[]
  {\label{fig:stabDiagN7}
    \includegraphics[width=0.48\linewidth]{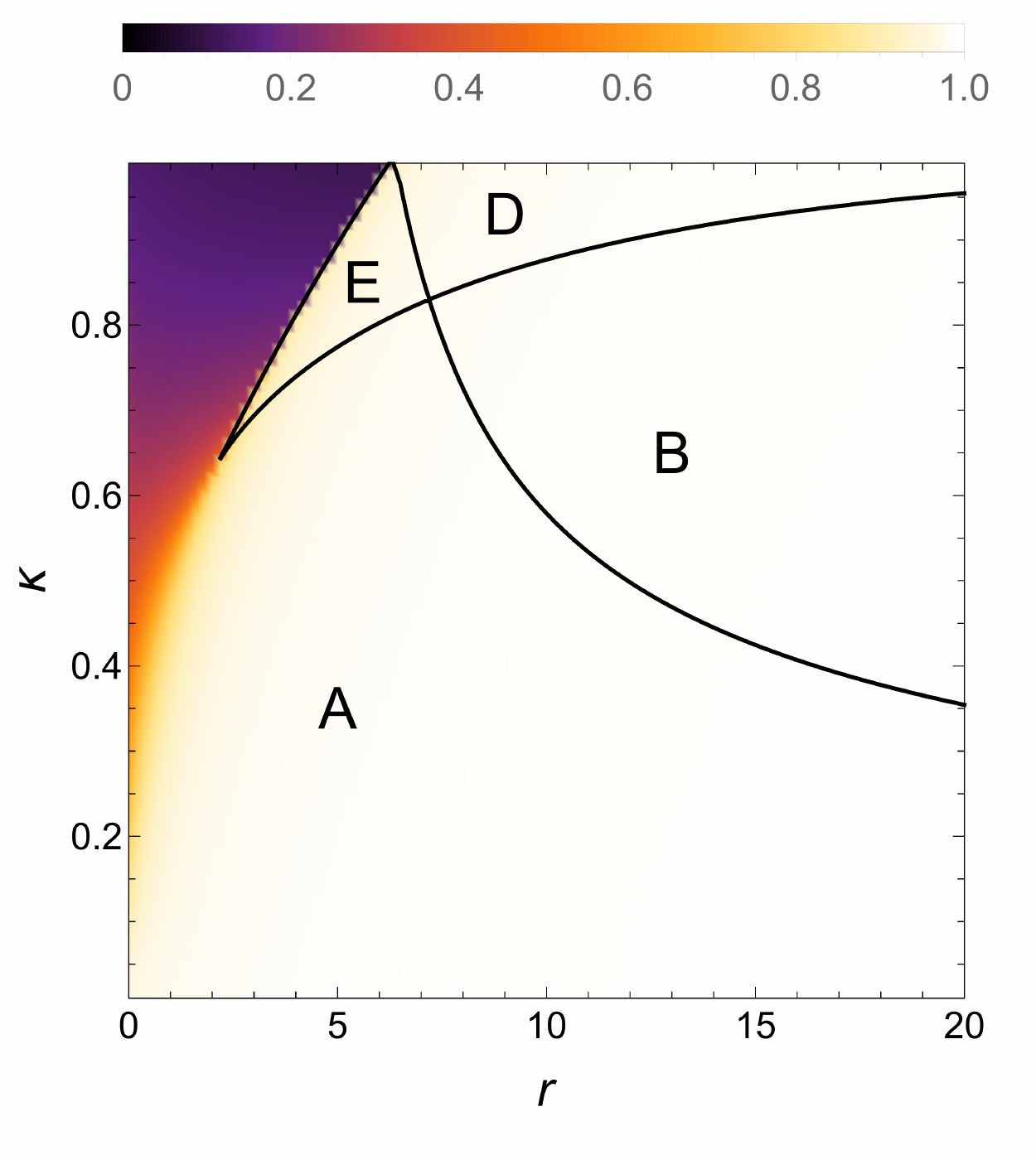}}\\
  \caption{{\footnotesize Stability diagrams for $v=5$ and
      $N \in \{4,5,6,7\}$, in panel \subref{fig:stabDiagN4},
      \subref{fig:stabDiagN5}, \subref{fig:stabDiagN6}, and
      \subref{fig:stabDiagN7}, respectively.  The area A indicates the
      systems phase with a single attractor in favor of the best
      option. Having an unique solution, in this area the system never
      converges for the selection of inferior options. The underlying
      density map shows the population size of the stable solution for
      the best option. In the dark area the population for the best
      option is not sufficient to reach a quorum to take a
      decision. For an increasing number of options, the dark area
      increases and low values of $r$ are not sufficient anymore to
      allow the swarm to take a decision for similar options (high
      $\kappa$). However, for sufficiently large values of $r$, the
      area A shifts towards higher values of $\kappa$. This effect is
      also shown in Figure~\ref{fig:stabilityRKN234} of the main
      text.}}
\label{fig:compareStabDiag}
\end{figure}

\begin{figure}[t!]
  \centering
 \subfigure[]
  {\label{fig:BifN4}
    \includegraphics[width=0.48\linewidth]{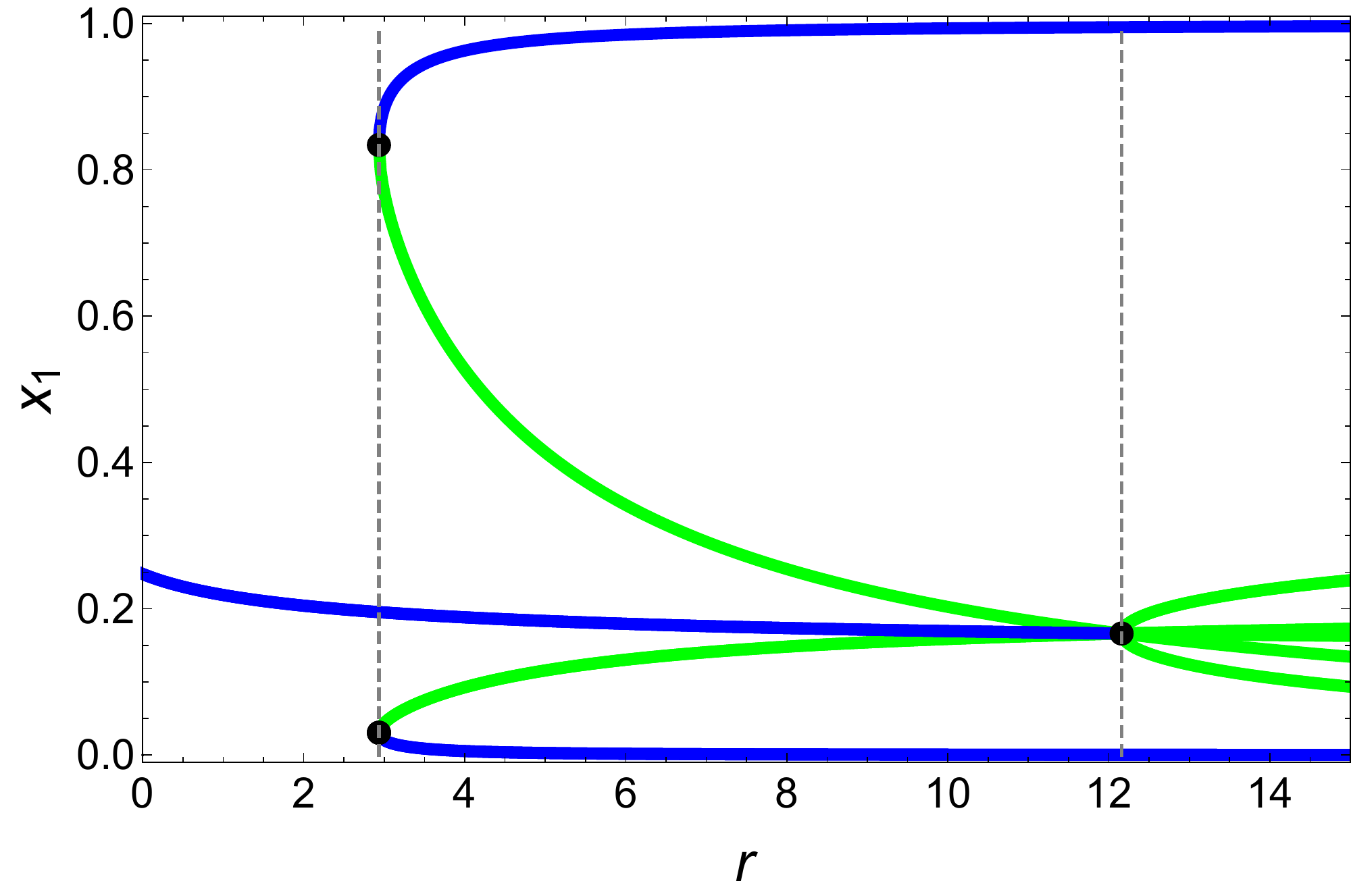}}
  \subfigure[]
  {\label{fig:BifN5}
    \includegraphics[width=0.48\linewidth]{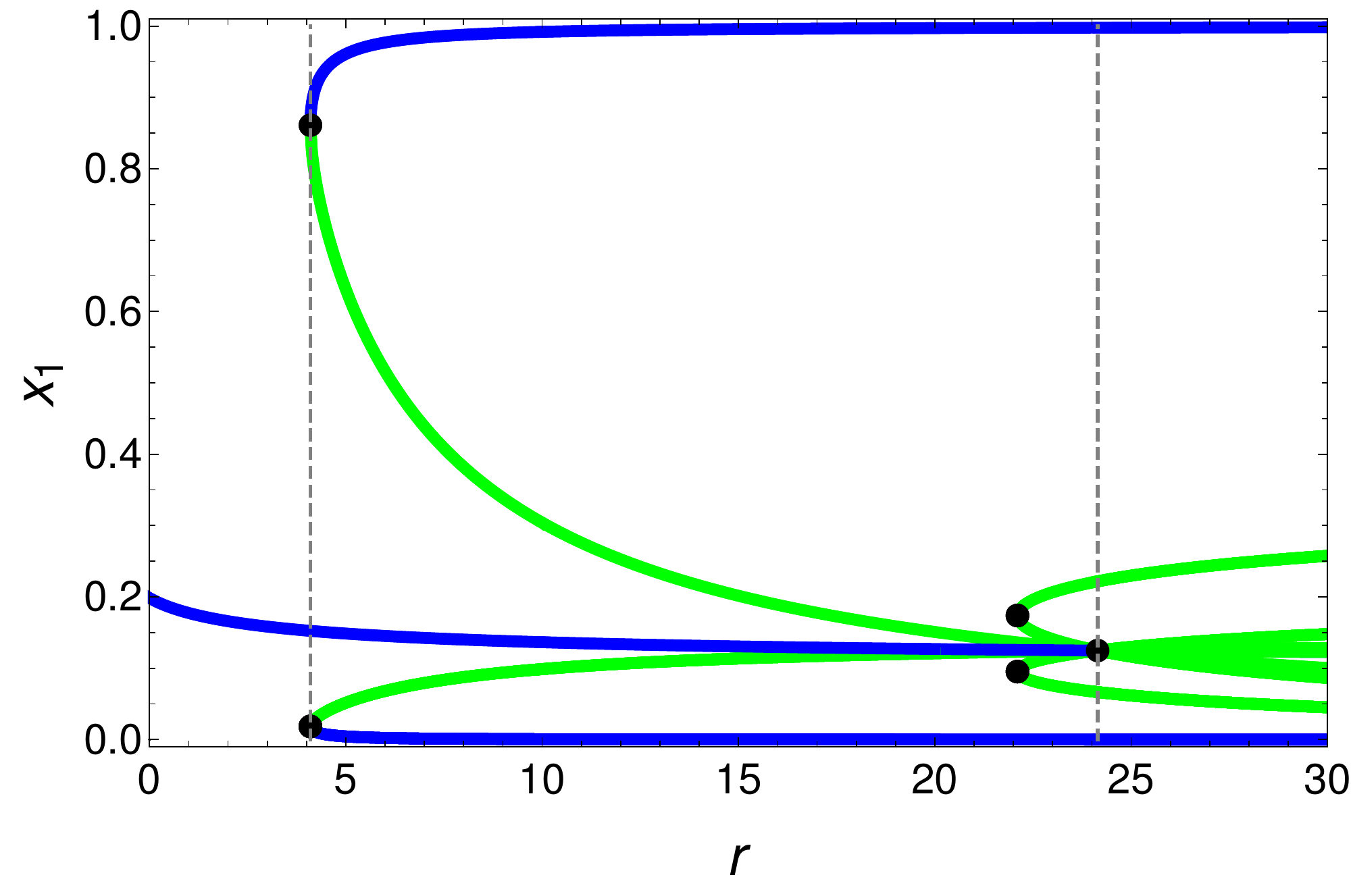}}\\
  \subfigure[]
  {\label{fig:BifN6}
    \includegraphics[width=0.48\linewidth]{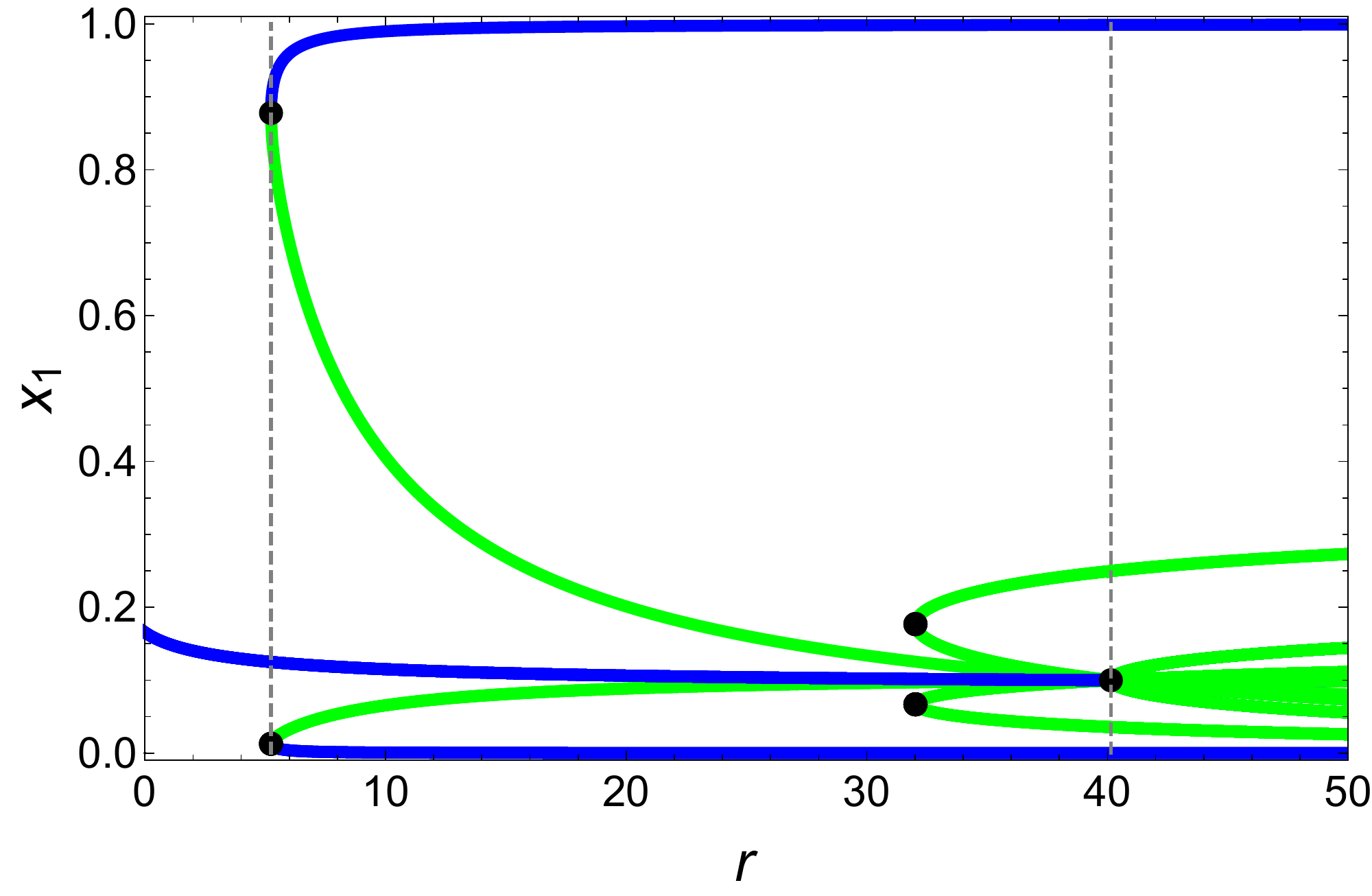}}
  \subfigure[]
  {\label{fig:BifN7}
    \includegraphics[width=0.48\linewidth]{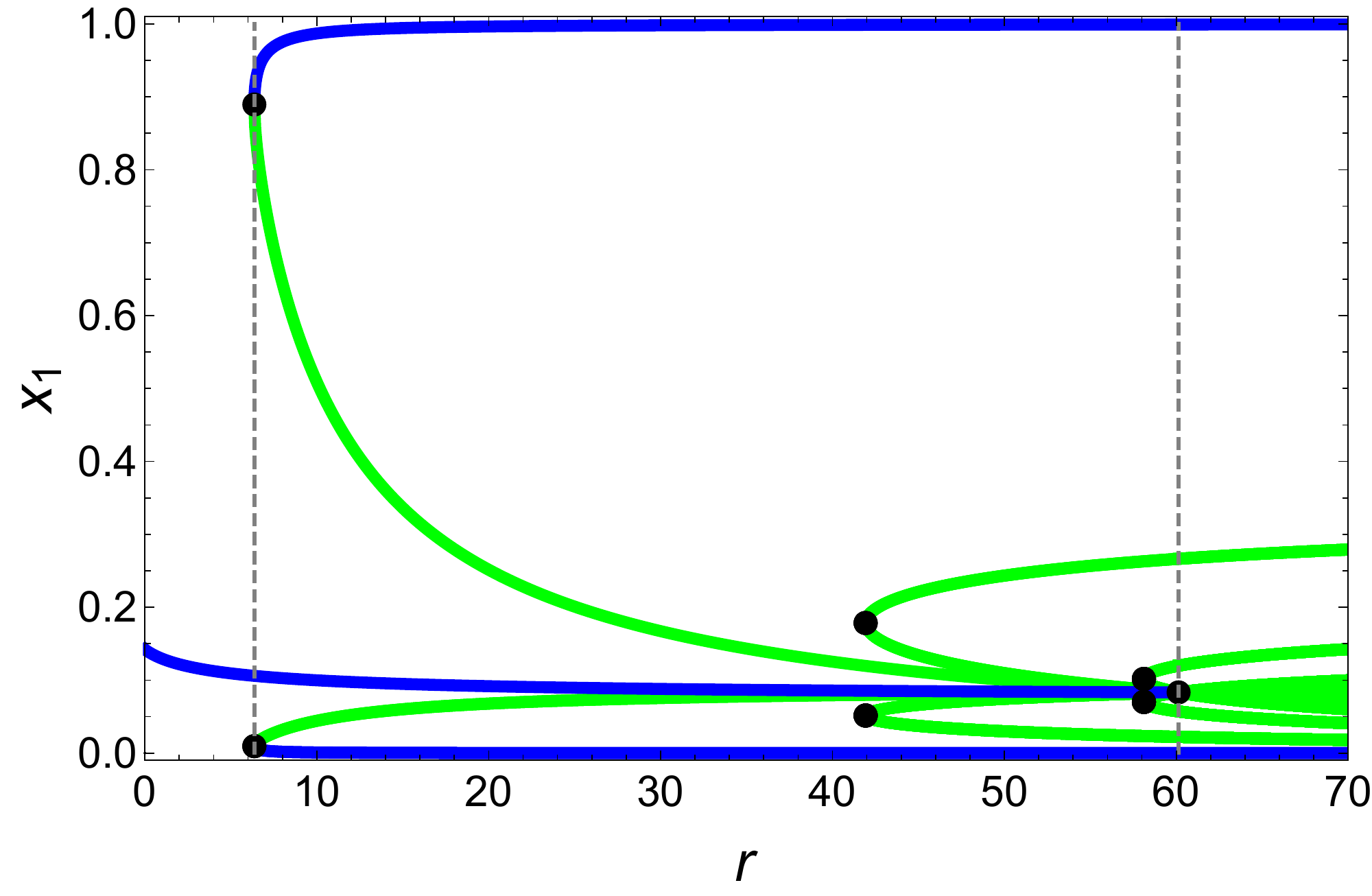}}\\
  \caption{Bifurcation diagrams of the complete system
    (Equation~\eqref{eq:fullSysBestOfN}) in the symmetric case ($v=5$)
    for number of options $N=4$ in panel \subref{fig:BifN4}, $N=5$ in
    panel \subref{fig:BifN5}, $N=6$ in panel \subref{fig:BifN6}, and
    $N=7$ in panel \subref{fig:BifN7}. Blue curves represent stable
    equilibria and green lines unstable saddle points. The vertical
    dashed lines are the bifurcation point predicted by the reduced
    system (Equation~\eqref{eq:bifurcations}). These points always
    precisely match with the bifurcation point of the complete
    system.}
\label{fig:bifurcationsManyNs}
\end{figure}

\begin{figure}[t!]
  \centering
  \subfigure[]
  {\label{fig:BifN4AS}
    \includegraphics[width=0.48\linewidth]{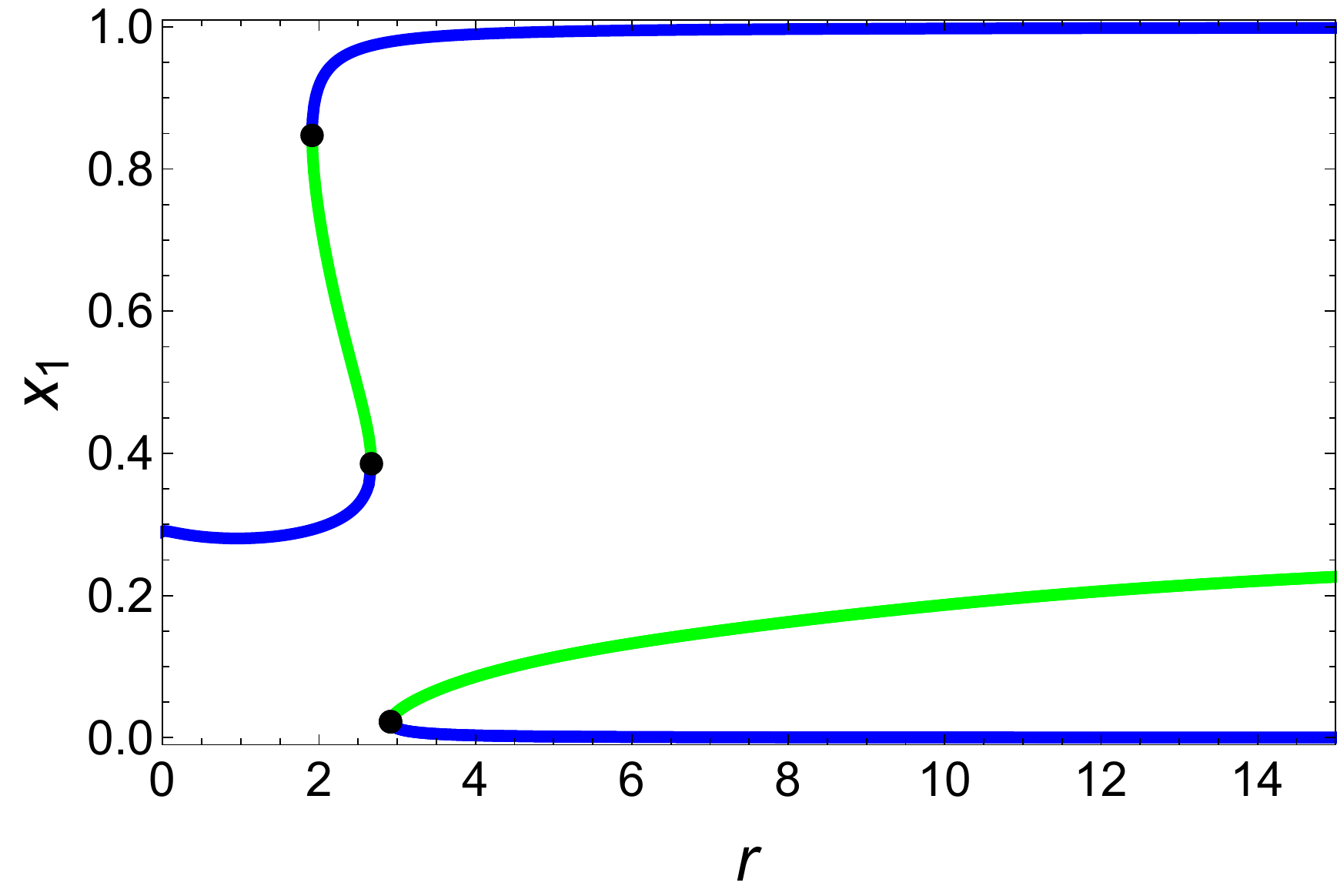}}
  \subfigure[]
  {\label{fig:BifN5AS}
    \includegraphics[width=0.48\linewidth]{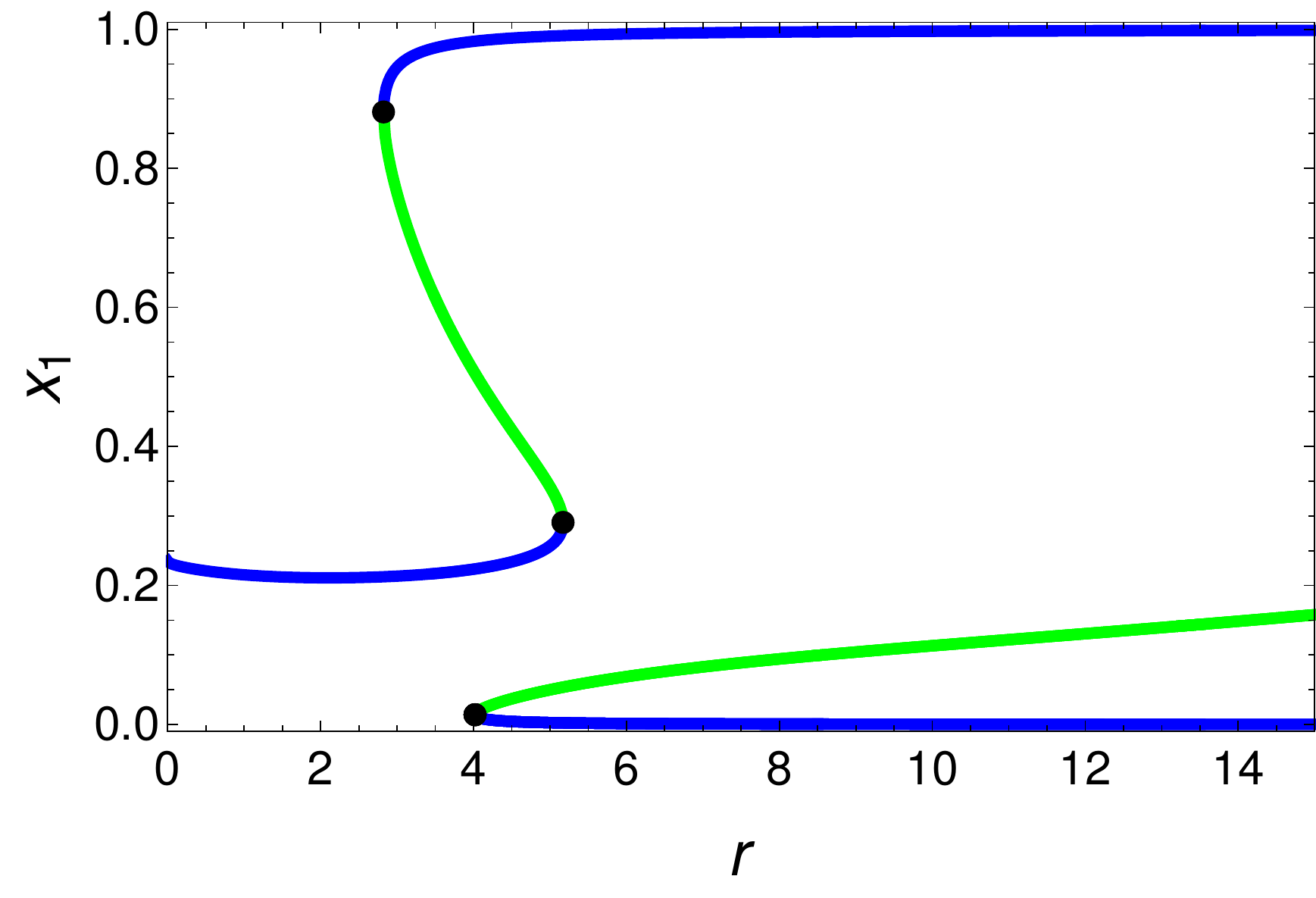}}\\
  \subfigure[] {\label{fig:BifN6AS}
    \includegraphics[width=0.48\linewidth]{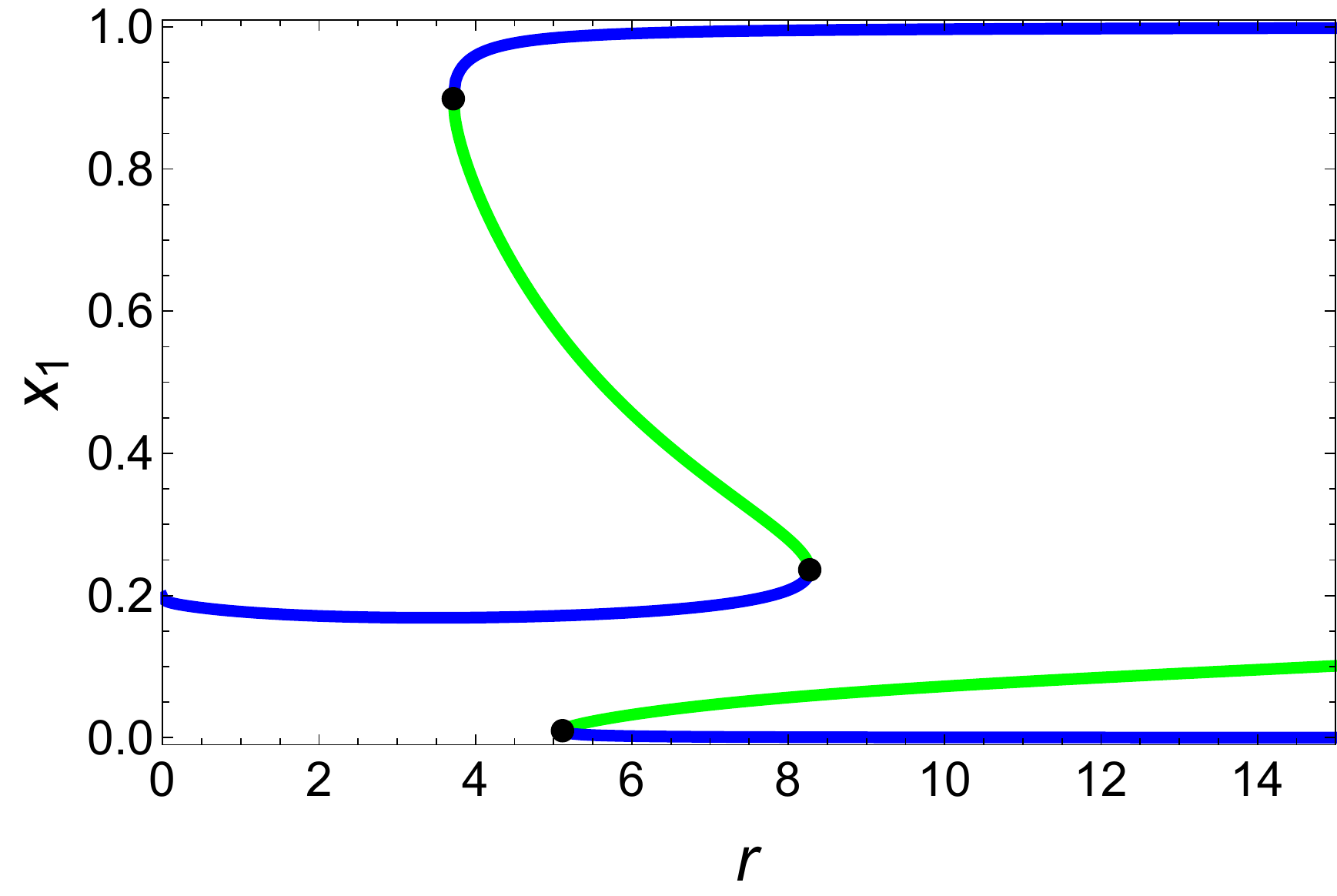}}
  \subfigure[] {\label{fig:BifN7AS}
     \includegraphics[width=0.48\linewidth]{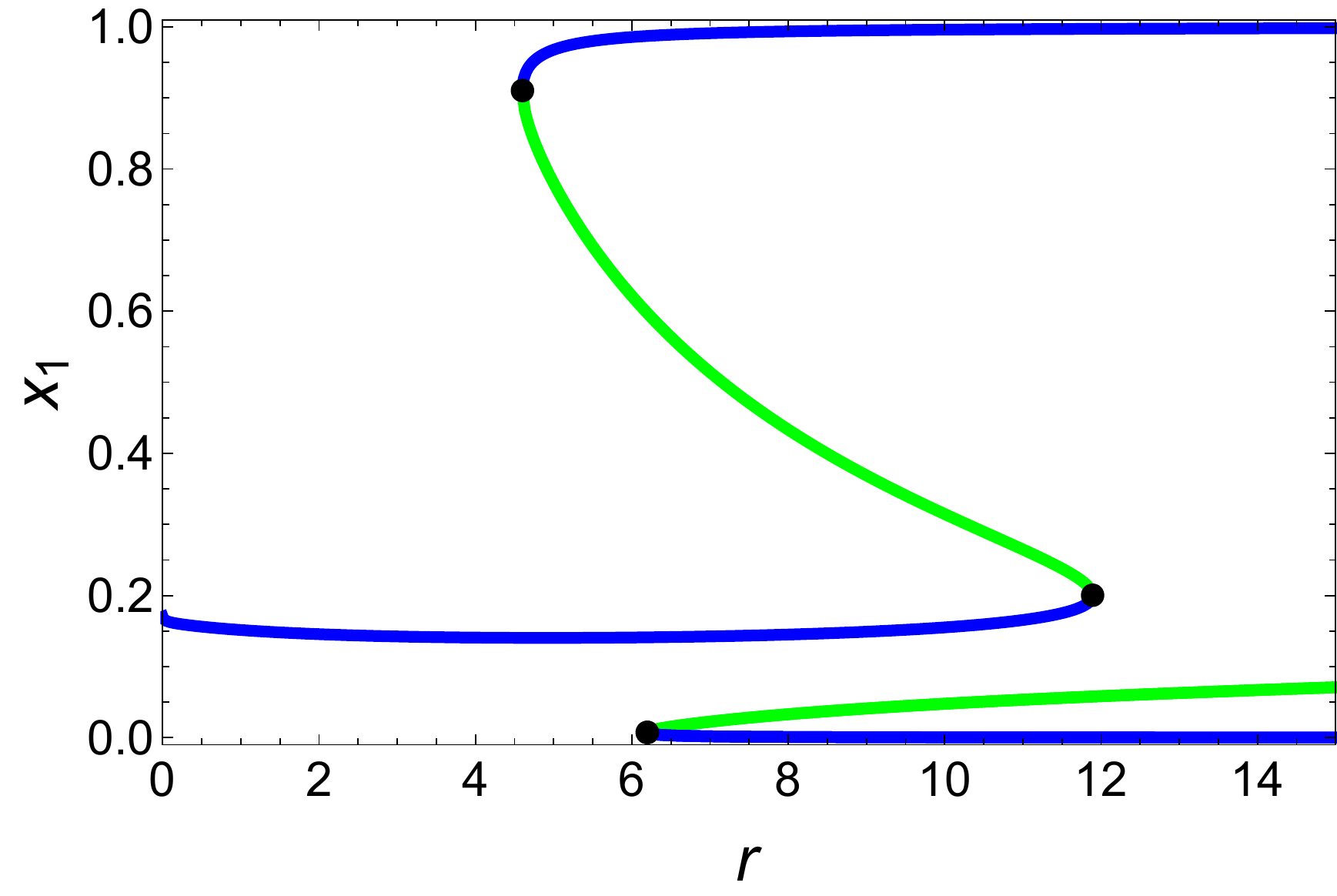}}
   \caption{Bifurcation diagrams of the complete system
     (Equation~\eqref{eq:fullSysBestOfN}) in the asymmetric case for
     number of options $N=4$ in panel \subref{fig:BifN4AS}, $N=5$ in
     panel \subref{fig:BifN5AS}, $N=6$ in panel \subref{fig:BifN6AS},
     and $N=7$ in panel \subref{fig:BifN7AS}. In all plots, the
     superior option's quality is $v_1=8$ while the inferior options'
     quality is $v_i=7.2,\; i \in [2,N]$, that is,
     $\kappa=v_i/v=0.9$. Blue curves represent stable equilibria and
     green lines unstable saddle points. Notice the increase of the
     range of values of $r$ in which the undecided state
     persists. Note also that the stable state at decision for the
     superior option appears earlier than the ones for the inferior
     alternatives. This supports a strategy to deal with the
     uncertainty in the decision-making scenario based on the gradual
     increase of $r$, which would initially bring the system into an
     indecision state and subsequently jump to the selection of the
     highest quality option.}
\label{fig:bifurcationsManyNsAsym}
\end{figure}

\vspace{0.7cm}
\textbf{Best of N.}
Figure~\ref{fig:compareStabDiag} shows the stability diagrams for
$N \in [4,7]$ with an underlaying density map showing the
population size for the best option. While area A corresponds to the
most favourable system phase, that is, there is one single attractor
with a bias for the superior option, however, in the dark shaded area
the population size is relatively low and might be not enough to reach
a decision quorum. The dark area increases with the number of options
$N$ and decreases with the difference in option's qualities (i.e.,
higher $\kappa$). Therefore, for similar options, higher values of $r$
(i.e., interactions) are necessary to let the swarm make a decision.

Additionally, we report the bifurcation diagram for $N \in [4,7]$ for
both the symmetric case (Figure~\ref{fig:bifurcationsManyNs})
and for the asymmetric case
(Figure~\ref{fig:bifurcationsManyNsAsym}).

\bibliography{Ref_Best_of_N}
\bibliographystyle{apsrev4-1}

\end{document}